\documentclass[preprint,12pt]{elsarticle}
\usepackage{amsmath,amssymb}
\usepackage{xcolor}
\usepackage{graphicx,graphics}
\usepackage{subcaption}
\usepackage{algorithm}
\usepackage{algpseudocode}
\usepackage{array}
\usepackage{xurl}
\newcolumntype{P}[1]{>{\centering\arraybackslash}p{#1}}
\usepackage[normalem]{ulem} 

\begin{document}
\setlength{\tabcolsep}{6pt}
\renewcommand{\arraystretch}{1.5}
\title{Indoor thermal comfort management: A Bayesian machine-learning approach to data denoising and dynamics prediction of HVAC systems}

\author[inst1]{J. Penuela}
\author[inst1]{S. Moghimian Hoosh}
\author[inst1]{I. Kamyshev}
\author[inst2]{A. Bischi}
\author[inst1]{H. Ouerdane}

\affiliation[inst1]{Center for Digital Engineering, Skolkovo Institute of Science and Technology, 30 Bolshoi Boulevard, Moscow 121205, Russia}
\affiliation[inst2]{University of Pisa, Department of Energy, Systems, Territory and Construction Engineering, Largo Lucio Lazzarino 1, Pisa, 56122, Italy}

\begin{abstract}
The optimal management of a building's microclimate to satisfy the occupants' needs and objectives in terms of comfort, energy efficiency, and costs is particularly challenging. This complexity arises from the non-linear, time-dependent interactions among all the variables of the control problem and the changing internal and external constraints. Focusing on the accurate modeling of the indoor temperature, we propose a data-driven approach to address this challenge. We account for thermal inertia, non-linear effects, small perturbations of the indoor climate dynamics caused by ventilation and weather variations, as well as for the stochastic nature of the control system due to the observed noise in the input signal. Since the prohibitive cost of quality data acquisition and processing limits the implementation of data-driven approaches for real-life problems, we applied a method that merges several Bayesian machine learning and deep learning architectures that are suitable for predicting complex system dynamics, while relaxing the dataset quality requirements. Our framework includes a built-in deep Kalman filter, which makes it deployable even with low-accuracy temperature sensors. It achieves state-of-the-art performance, best performing with a 150-minute prediction horizon with an RMSE of 0.2455, an MAE of 0.162, and an $R^2$ of 0.926. The model's performance remains consistent even when exposed to highly noisy data. Finally, we show how our approach can be extended to other applications including demand response event duration prediction and equipment failure detection. 
\end{abstract}
\date{\today}

\begin{keyword}
Indoor thermal comfort \sep HVAC systems \sep Bayesian machine learning\sep data analytics\sep synthetic data\sep energy in buildings
\end{keyword}

\maketitle

\section{Introduction}
Economic growth and improvement of quality of life correlate with increased energy consumption \cite{khan2021dynamic,QoL2018}, especially in developing countries where the industrial base is expanding \cite{QoL2018,IEA2024}. Statistics published in the International Energy Agency report ``World Energy Outlook 2024'' \cite{IEA2024}, show that energy demand has increased by 1.4\% per year on average since 2010 and that some notable trends require particular attention. First, the yearly electricity demand growth rate is twice that of the overall energy demand, indicating a strong shift toward electrification, including for cooling and heating applications. Second, this growth is expected to slow to 0.5\% per year due to the global shift toward a service economy, which is less energy-intensive than a heavy industry economy. Finally, the main driver of demand growth is expected to be the energy consumption in residential and commercial floor areas, which will be multiplied by a factor of 1.5 by 2050 \cite{IEA2024}. In developed economies, a significant portion of household energy use is dedicated to thermal comfort. For instance, in the European Union, 80\% of household energy consumption is used for heating, cooling and hot water \cite{directorate-general}, which highlights the importance of thermal comfort in developed economies. As a result, we can expect the global demand to follow the current trends of developed countries. According to the \cite{IEA2024}, the cooling energy demand will increase worldwide to 1200 TWh by 2035, an amount comparable to the electrical energy consumed in the entire Middle East in 2024. These trends result in rising energy costs \cite{Vulnerable2023,BAYERAMADESSA2024118210} and air pollution \cite{khan2021dynamic, Vulnerable2023} with adverse effects on economies and public health. In this context, energy efficiency is essential for improving quality of life while mitigating the negative effects of globally growing electricity consumption. This is particularly applicable to indoor microclimate control where thermal comfort leads to constantly growing consumption due to increased living standards in developing countries \cite{IEA2024}. 

Indoor thermal comfort is based on standards that define an adequate range of temperatures in the built environment (residence, office, shopping mall, factory, etc.) depending on their use and the habits of occupants \cite{Ortiz2017,ASHRAE}. Hence, the acceptable temperature range around an optimal temperature can vary from several degrees Celsius ($^{\circ}$C) to a much narrower range, e.g., $\pm 1^{\circ}$C, depending on the context. It must be noted that each scenario imposes different requirements on HVAC control systems, and random events caused by occupant behavior or faulty sensors can further affect system performance. Concretely, in residences, the optimal temperature window recommended by the World Health Organization is in the 18-24$^{\circ}$C \cite{Health_thermal_confort2012}. While maintaining thermal comfort in high-income households with healthy occupants poses no problem, certain factors such as age, health issues or poverty may impose severe constraints on how the room temperature is set or even make achieving thermal comfort unfeasible\cite{ZHANG2024111137}. For instance,  elderly people suffering from hypothyroidism often feel cold even at 24°C, since their bodies produce less heat than normal, and medication cannot fully solve this problem \cite{Romero-Ibarguengoitia2024-da}. Another example, economic constraints necessitate low-income households lowering the temperature to 18–19$^{\circ}$C \cite{Vulnerable2023, Health_thermal_confort2012}. Note that, the combination of constraining factors, as in the case of maintaining thermal comfort for the elderly with hypothyroidism in low-income households, can lead to unfeasible solutions.

In industrial settings, maintaining an appropriate temperature range is not just about comfort but also about ensuring product quality and process efficiency. Thus, depending on the industrial processes, temperature constraints can be either relaxed or strict. For instance, in industrial greenhouses, the production of tomatoes requires an optimal temperature close to 25$^{\circ}$C during day time \cite{Heat_stress1998}. As a general rule, the thermal comfort window for plants is approximately $\pm$ 6$^{\circ}$C around the optimal temperature \cite{Review_dynamic_VF2024}. This means that, for most stages of tomato plant and fruit development, temperatures up to 29$^{\circ}$C remains within the comfort zone. However, during pollen production, a temperature of 29$^{\circ}$C is the threshold when the pollen from the tomato plants becomes completely sterile. Given the complexity of indoor microclimates, an advanced heating ventilation and air-conditioning (HVAC) control system that can assess constraints and optimize both energy consumption and thermal comfort is a must-have for households and industries.

A common way to classify control approaches to regulate HVAC systems is by dividing them into two categories: model-free and model-based \cite{model_vs_modelfree2023}. Among model-free methods, rule-based control is widely used due to its simplicity but it cannot manage the complex dynamics of indoor microclimate, which makes it not optimal \cite{model_vs_modelfree2023}. Many recent studies have focused on reinforcement learning (RL), a model-free machine learning (ML) method which can be used for controlling the complex systems \cite{RL, buşon_2018, yang_2020}, including indoor environment microclimate \cite{Value_based,hosseinloo2020datadriven}. RL algorithms are often classified into value-based \cite{Value_based} and policy-based methods \cite{Policy_based}. To be applicable, both types of RL methods require only the Markov property (the next state depends only on the current state and action). Further, RL methods converge in highly unstable ways due to sampling variance, which requires considerable effort to tune the model's hyperparameters \cite{HPO,ZHOU2025112663}. In practice, hybrid approaches to relax constraints on the Markov property and reduce sampling variance are widely used including, data-driven methods or simulations of the environment to accelerate online learning and reduce data variance \cite{DDPG,ZHOU2025112663}, thereby improving the stability of RL algorithms. The main disadvantage of RL based approaches is that either its proper implementation necessitates a large amount of data or it becomes computationally costly.

Turning to model-based control approaches, linear model predictive control falls short in efficiency to describe complex nonlinear dynamics \cite{model_vs_modelfree2023,FRAGUELADIAZ2025111924,MPC_our}. Among the best performing non-linear models, nonlinear model predictive control (NMPC), and machine learning (ML) methods are the most promising approaches in control systems, particularly thermal comfort control in buildings \cite{Seyedzadeh2018,MOCANU201691}. To date, several ML methods, including support vector machines, decision trees, statistical algorithms, Bayesian networks, and Recurrent neural networks have been applied to different problems in indoor thermal control, showcasing the good performance, with low prediction errors \cite{AMASYALI20181192,Real_time_prediction2018}.  However, like all data-driven approaches, ML methods require a significant amount of real-world data. Furthermore, as these approaches are often fine-tuned to the specific data for training, their ability to generalize to new data from similar buildings typically requires additional exposure to the new data.

In contrast to the ``black box'' approach that ML represents, microclimate dynamics can be modeled based on the physical principles of mass and energy conservation which is known as ``white box''. Examples of such models include commercial software such as, e.g., TRNSyS \cite{TRNSYS} and EnergyPlus \cite{energyplus}, which allow for detailed design of buildings and microclimate dynamics simulation \cite{8282965}. Such models can provide accurate enough dynamics that can be utilized for controller design. However, the design process is time-consuming and requires significant modeling efforts and expert know-how, which precludes the widespread use of the software. Moreover, changes in buildings and systems with time should inevitably lead to an update of the model, which complicates the controller support. Another point of concern is the presence of stochastic and uncertainty effects influencing the controller behavior. As this cannot be accounted for by this type of model, the quality of the predictions is reduced \cite{model_vs_modelfree2023}.

Recent research suggests that \textit{hybrid} approaches combining data-driven models with physics-based models can address these limitations \cite{model_vs_modelfree2023}. This combination is often referred to as a ``grey box'' approach where both types of models can be merged into a single larger model. In fact, the physics-based model discards all unrealistic predictions, and hence acts as a very strong regularizer, stabilizing learning and prediction processes of the data-driven model. Allowing the data-driven model to find unknown parameters of the physics-based model and add observed information from the physical processes beyond the assumptions and limitations required for the physical model to be computable. This approach can significantly improve the generalization capabilities of the model, while keeping the advantages of data driven approaches \cite{model_vs_modelfree2023}. Bayesian deep learning has advantages over the other grey box methods as it allows to embed prior knowledge into the model while also allowing to handle the inherently stochastic and noisy real world data efficiently, which is reflected in higher data efficiency and relaxed dataset requirements \cite{LINKA2022115346}.

Overall, in spite of significant progress in the field of optimal indoor microclimate control in the recent years, the following shortcomings remain to be addressed:

\begin{itemize}
    \item Effective models at short horizons tend to under-perform at long horizons and conversely, which limits the scope of their applications because of varying reliability for different prediction horizons.
    \item  High-performance models like digital twins or specialized neural networks lack generalization capabilities and require expert work (as for the initial training) to fine tune the model using large amounts of quality data \cite{Arango2023QuickTuneQL}. 
    \item Lack of modularity as there is only a choice between all-in-one control predictive architecture and stand-alone pieces (model, filter, control logic, etc.) The former limits the applicability of developed tools, while the latter results in overcomplicated implementations.
    \item Lack of robustness to unobservable non-linear variables including measurement noise and thermal inertia.
\end{itemize}

To address these limitations, we proposed a modular, and hybrid model based on the Bayesian neural networks for indoor temperature prediction. The model is rooted in the combination of Bayes' theorem and computational models that allows incorporating the prior knowledge of the system into the model, which reduces data requirements for successful fine tuning. 

In this work, we modeled a single room building in a ``warm summer humid continental'' climate, with a realistic weather, and HVAC response to generate a synthetic dataset. Later, we implemented a Bayesian neural network based on the well-studied neural networks for denoising and indoor temperature prediction at different prediction horizons. We addressed the complexity and scalability issues by combining deep learning methods \cite{Goodfellow-et-al-2016} and variational inference \cite{ELBO, igea_chatzis_cicirello_2022}. Our proposed model is scored using the standard metrics for regression, including, root mean squared error (RMSE), mean absolute error (MAE), and the coefficient of determination $R^2$. We scored our model for different prediction horizons and included a fidelity band method, exploring the trade-offs in relaxation in few-minute violations of the fidelity band. This is well-suited for tasks that look for divergence from measurements rather than for short-lived events. Also, we present state-of-the-art benchmarks from the scientific literature. The main contributions of this paper are as follows:

\begin{itemize}

    \item The modular design of this model allows the different components of our model such as denoising, or prediction to be used independently. 
    
    \item To address the memory effects in energy systems, and to capture long-term dependencies, we propose using a recurrent neural network with a gated recurrent unit, which is well suited to problems involving memory effects \cite{DARN}.

    \item We implement a deep Kalman filter with self-calibrating capabilities, using a one dimensional convolutional neural network which is robust to noise levels beyond those used during the training process based on \cite{DKF}. 
    
    \item Our proposed training method is based on variational inference, which is one of the most efficient probabilistic approaches to machine learning \cite{igea_chatzis_cicirello_2022}. Furthermore, we achieve probabilistic modeling at the cost of deterministic modeling by leveraging the intrinsic distribution of measurement noise.

    \item We achieved RMSE of 0.2455, MAE of 0.162, and $R^2$ of 0.926 for a 150-minute prediction horizon. Compared to prior studies, the model achieves lower RMSE (0.2–0.27 $^{\circ}$C) than existing HVAC prediction models, which typically show RMSE 0.6 $^{\circ}$C. Results indicate that the performance remains stable across different prediction horizons (30–2400 minutes).

\end{itemize}  

The article is organized as follows. In section II, we address the modeling and simulation of the indoor temperature dynamics involving the generation of synthetic data on the one hand, and its probabilistic description on the other hand. In section III, we present the proposed Bayesian ML architecture, including descriptions of the training method and the loss function. Section IV is devoted to the obtained results, including performance metrics, and further application of our trained model to HVAC control: fine-tuning, examples of use, including demand response participation with thermal loads and equipment fault detection. Section V ends the article with our conclusions.

\section{Methodology}

\subsection{Indoor microclimate model and simulation}
\label{simulation}
In this work, using the following assumptions, we simulate a single thermal zone (single room) building, which allows for a clear interpretation of the results accounting for the complexity of the indoor environment subjected to the weather as an external constraint: 
 \begin{itemize}
    \item The room is sufficiently homogeneous to use a lumped simulation method. In practice, this reflects the fact that many households do not yet have modern, programmable thermostats, as is the case for 70\% of American households \cite{PEFFER20112529}. This implies that a single sensor provides sufficient information, but it does not. 
    \item We consider two parameters to characterize thermal comfort: the observed (or measured) temperature in the room $T_{\rm obs}$ and the ventilation in the room, $V$, which regulates the concentration of carbon dioxide. This constitutes a microclimate state. 
    \item The ventilation works in a linear regime: when the ventilation is in the OFF (resp. ON) position, the concentration of CO$_2$ in the room grows (resp. decreases) linearly with time, with a coefficient $c_{\rm off}$ (resp. $c_{\rm on}$).
    \item The outdoor temperature, $T_{\rm out}$, is an external observable state variable, which influences the control system behavior.
    \item Heating, ventilation and cooling are controlled. The control space is defined for all devices by binary state variables with values 1 or 0 characterizing the ON or OFF state: $a_{\rm h} = \{0, 1\}$, $a_{\rm vent} = \{0, 1\}$ and $a_{\rm ac} = \{0, 1\}$.
    \item Thermal inertia is characterized by three unobserved (i.e. not measured by sensors) parameters: the average temperature of the room's walls, floor and ceiling, $T_{\rm w}$, the temperature of the air near the heater $T_{\rm h}$, the air conditioner coils temperature $T_{\rm ac}$. Note that as cooling relies on smaller temperature gradients and forced convection, we assume that the cooling system inertia is insignificant; hence, $T_{\rm ac}$ has comparatively a smaller value than $T_{\rm w}$ and $T_{\rm h}$, and does not influence the room temperature when the air conditioner is off.
    \item The observed temperature can be written as: $T_{\rm obs} = T + \epsilon$, where $T$ is the true average temperature in the room at time $t$, and $\epsilon$ is the measurement noise which is assumed to be normally distributed with a zero mean $\epsilon \sim \mathcal{N}(0, \sigma)$. Both $T$  and $\epsilon$ are unobservable variables -- only their sum is observable.
    \item We selected a sampling frequency of one observation per minute as it is the highest available resolution for weather forecasts \cite{openweathermap.org_2012}. Also, we show that this granularity level may well apply to problems such as demand response and fault detection as discussed in Sec. \ref{sec:DR} and \ref{sec:Faults}. 
    \item The outside temperature $T_{\rm out}$ is assumed to be observed with a high precision $\epsilon_{\rm out}$ such that $\epsilon \gg \epsilon_{\rm out}$. We consider changes during a 24-hour period mimicking the effects of solar radiation on the weather by sampling from a cosine function \cite{SPROUL2017292}.
    
\end{itemize}

The system of equations modelling the dynamics of the simulated micro-climate accounting for the (simulated) weather, based on the assumptions above, thus reads: 
\begin{eqnarray}
\label{eq:Tobs}
T_{\rm obs}^{\rm t+1} &=& T^{\rm t+1} + \epsilon\\
\nonumber
T^{\rm t+1} &=& T^{\rm t}+ (T_{\rm h}^{\rm t} - T^{\rm t})f(T_{\rm h}^{\rm t}, T^{\rm t}) + (T_{\rm w}^{\rm t} - T^{\rm t})g(T_{\rm w}^{\rm t}, T^{\rm t})\\ 
&+& a_{\rm vent}^{\rm t} (T_{\rm out}^{\rm t}- T^{\rm t})\phi(T_{\rm out}^{\rm t}, T^{\rm t}) + a_{\rm ac}^{\rm t}  (T_{\rm ac}^{\rm t}- T^{\rm t}) h(T_{\rm ac}^{\rm t},T^{\rm t})\\
T_{\rm h}^{\rm t+1} &=& T_{\rm h}^{\rm t} + f_{\rm h}(T_{\rm h}^{\rm t}, a_{\rm h}^{\rm t}) + (T^{\rm t} - T_{\rm h}^{\rm t})g_{\rm h}(T^{\rm t}, T_{\rm h}^{\rm t})\\
T_{\rm ac}^{\rm t+1} &=& T_{\rm ac}^{\rm t} + f_{\rm ac}(T_{\rm ac}^{\rm t},T^{\rm t})\\
T_{\rm w}^{\rm t+1} &=& T_{\rm w}^{\rm t}+(T^{\rm t} - T_{\rm w}^{\rm t})f_{\rm w}(T^{\rm t}, T_{\rm w}^{\rm t}) + (T_{\rm out}^{\rm t} - T_{\rm w}^{\rm t})g_{\rm w}(T_{\rm out}^{\rm t}, T_{\rm w}^{\rm t})\\
V^{\rm t+1} &=& V^{\rm t} + a_{\rm vent}^{\rm t} \rm c_{\rm on} - (1 - a_{\rm vent}^{\rm t})\rm c_{\rm off}\\
\label{eq:Tout}
T_{\rm out}^{\rm t+1} &=&T_{\rm out}^{\rm t} + f_{\rm out}(T_{\rm out}^{\rm 1},T_{\rm out}^{\rm t}) + h_{\rm out}(\rm t) + g_{\rm out} (X \sim \mathcal{N}(0,1))
\end{eqnarray}

\noindent where $f$, $g$, $\phi$, $h$, $f_{\rm h}$, $g_{\rm h}$, $f_{\rm ac}$, $f_{\rm w}$, $g_{\rm w}$, $c_{\rm off}$, $c_{\rm on}$, $f_{\rm out}$, $h_{\rm out}$ and $g_{\rm out}$ are non-negative functions and parameters characterizing non-linear effects on the time evolution of the indoor temperature and CO$_2$ concentration (See \ref{Appendix:A} for additional details). They are selected to simulate a single-room building that can maintain a set temperature of 25$^{\circ}$C with an outside temperature range from $-$30$^{\circ}$C to 35$^{\circ}$C which covers the possible temperature range in a ``warm summer humid continental'' climate. Note that in this model, the function $g_{out}$ mimics the combined effects of environmental variables changes (humidity, wind direction and speed, and cloudiness denoted with the random variable $X\sim \mathcal{N}(0,1)$ in Eq.~\eqref{eq:Tout}) on the short-term variations of $T_{\rm out}$. Also, as the variations in daily temperature usually are not extreme (up to 20$^{\circ}$C between the minimum and maximum temperatures throughout the day), we add a weak penalization constraint to the model to ensure that $T_{\rm out}$ is bounded.

\begin{figure}[ht]
	\includegraphics[width=\linewidth]{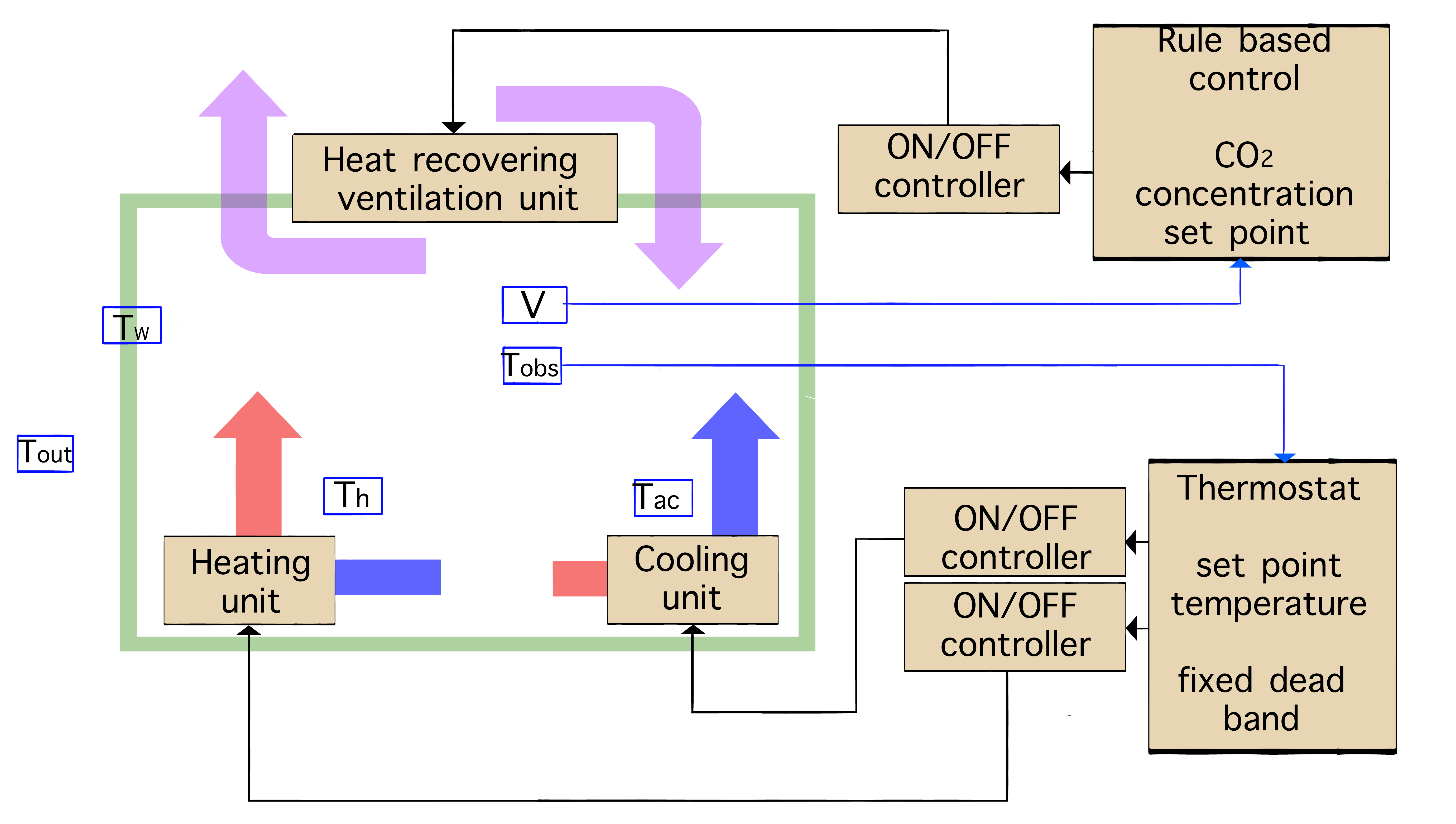}
	\caption{Graphical representation of single thermal zone test case used for synthetic data generation.}
	\label{fig:testcase_TRNSyS}
\end{figure}

\begin{figure}[ht]
	\includegraphics[width=\linewidth]{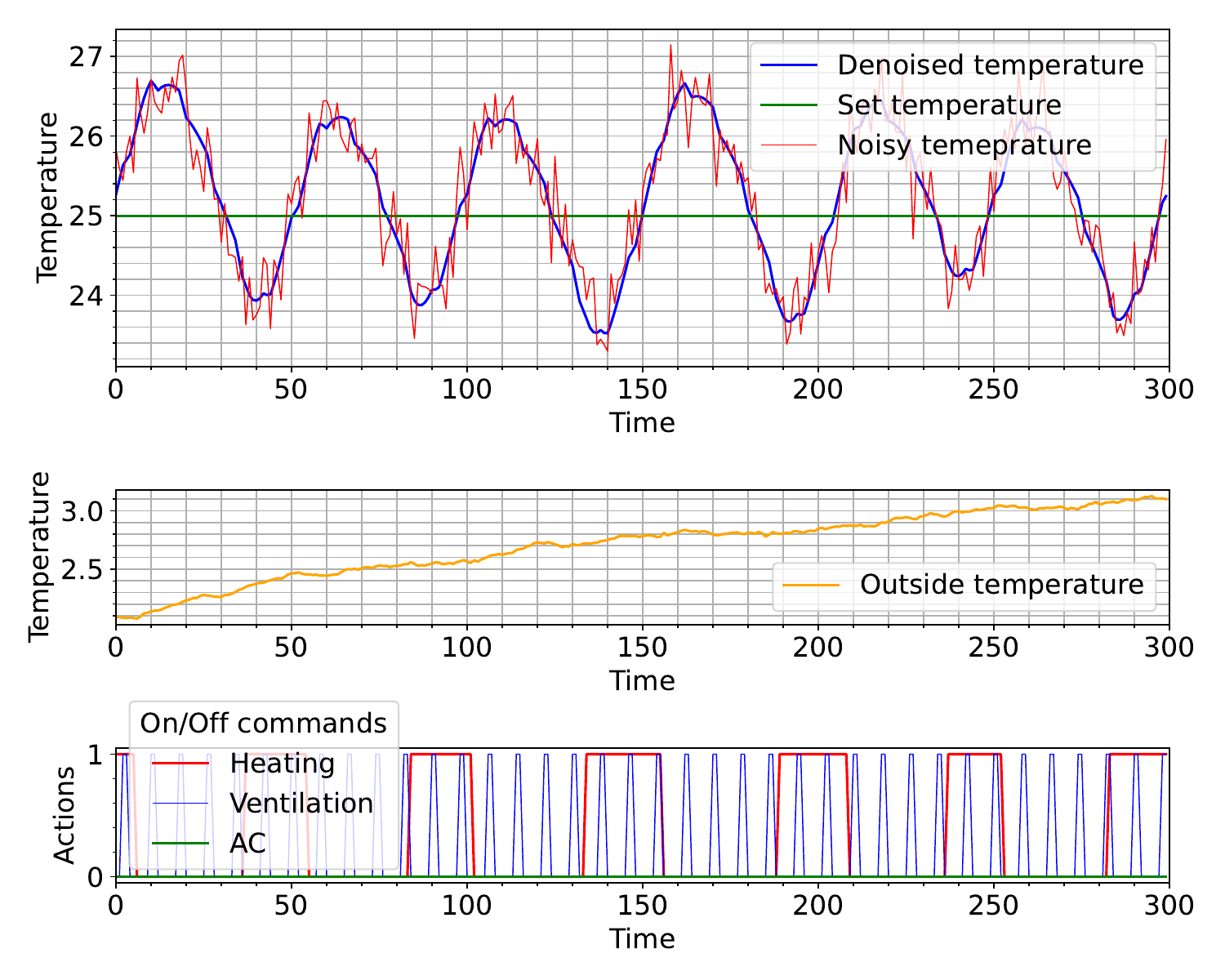}
	\caption{Example of synthetically generated data using the model presented in Section \ref{simulation}.}
	\label{fig:synthetic_data}
\end{figure}

Owing to its simplicity, the model can simulate a wide class of dynamics corresponding to different building types and climates. A graphic representation of the model is presented in Fig. \ref{fig:testcase_TRNSyS}. We use this model to generate a synthetic dataset for the training and validation stages of the neural network proposed in section \ref{sec:BML_model}. A sample of the dataset is presented in Fig. \ref{fig:synthetic_data}. In sections \ref{sec:DR} and \ref{sec:Faults} respectively, modified versions of this simulation approach are used to generate synthetic data for the proposed applications: demand response using the thermal-load and HVAC system fault detection. 

\subsection{Indoor microclimate probabilistic description}
\label{sec:BML_model}
Our approach follows a previous work on deep Kalman filter for climate control \cite{DeepKalmanfilterforclimatecontrol}, which is based on deep Kalman filter \cite{DKF} and variational autoencoder \cite{VAE} approaches. The generated synthetic data can be approximated by probabilistic modeling, allowing to pass prior knowledge of the microclimate state to the Bayesian machine learning model, hence making it possible to reconstruct the different unobserved physical parameters of our simulated test building with a  Bayesian model \cite{BML}. We use historical data to define all unknown parameters in the model by using our prior knowledge of the modeled measurement noise and thermal inertia. The observed parameters are the microclimate state variables, the external state variables and the control system response state, as defined in section \ref{simulation}. 

The probability of observing the noisy temperature $T_{\rm obs}$ with a given noiseless temperature $T$ at any time $t$ follows a normal distribution. The uncertainty between the measured (noisy) temperature and the actual (noiseless) temperature reflects the influence of external noise, which we assume is some white Gaussian noise stationary in time. Therefore, we can describe the conditional probability density function (pdf) of observing $T_{\rm obs}^{\rm t}$ given $T^{\rm t}$ as follows:
\begin{eqnarray}
P[T_{\rm obs}^{\rm t}|T^{\rm t}] = \frac{1}{\sqrt{2\pi}\sigma}\exp\left(-\frac{(T_{\rm obs}^{\rm t} - T^{\rm t})^2}{2\sigma^2}\right)
\label{eq:Pnoise}
\end{eqnarray}
 In our simulation, we modelled thermal inertia as a memory-less system involving unobserved  variables $T_{\rm h}$ and $T_{\rm w}$. Here to take into account thermal inertia, we consider dynamics with memory (non-Markovian dynamics) of $T$ instead of memory-less dynamics in expanded space (with unobserved  variables $T_{\rm h}$ and $T_{\rm w}$). The pdf of $T^{\rm t}$ given the history of states from observation $1$ to observation $\rm t-1$ of our observed variables $T^{\rm t-1:1}, a^{\rm t-1:1}, T_{\rm out}^{\rm t-1:1}$ is described as follows:
\begin{eqnarray}
&&P_\theta[T^{\rm t}|T^{\rm t-1:1}, a^{\rm t-1:1}, T_{\rm out}^{\rm t-1:1}] = \frac{1}{\sqrt{2\pi}s}\exp\left(-\frac{(T^{\rm t} - \mu_\theta^{\rm t})^2}{2s^2}\right)
\label{es:Ppred}
\end{eqnarray}
where $\mu_\theta^{\rm t}=\mu_\theta(T^{\rm t-1:1}, a^{\rm t-1:1}, T_{\rm out}^{\rm t-1:1})$ reflects the system memory and is described using a parametric family of functions. $\mu_\theta$ returns the expected value of the next temperature of the indoor air. Note that $\mu_\theta$ depends on all previous values of indoor temperatures, outside temperatures, and history of control signals. Also, notice that we use $t:1$ in our notation for the range of observations $\{t, t-1, \dots, 1\}$. We assume that the room parameters change very slowly. Hence,  they are stationary in time. We can exploit this feature in different applications, including fault detection, which is discussed in section \ref{sec:Results}. The most natural choice to approximate the family of functions $\mu_\theta$ is a recurrent neural network (RNN) \cite{RNN}. The particular form of this RNN is discussed in Section \ref{RNNarch}. 

The most common approach to find parameters $\theta$  of a statistical model (i.e., a suitable parametric family of functions) $P_\theta$ is the maximum likelihood method (MLE) \cite{MLE}. In MLE we aim to discover the set of parameters $\hat{\theta}$ called the maximum likelihood estimator, searching among the parameters $\theta$ over the parameter space $\boldsymbol{\theta}$. The maximum likelihood estimator $\hat{\theta}$ maximizes the likelihood function $\mathcal{L}_{\theta}$, making the observed data most probable under the assumed statistical model. For practical reasons, we use log-likelihood instead of likelihood:
\begin{eqnarray}
\nonumber
{\cal L_{\theta}} & = & \log P_\theta[T_{\rm obs}^{\rm t:1}|T^{\rm t - 1:1}, a^{\rm t-1:1}, T_{\rm out}^{\rm t-1:1}]\\
&=& \log \int\!\! dT^{\rm 1},\dots,dT^{\rm t}\prod_{\tau=2}^t P[T_{\rm obs}^{\rm \tau}| T^{\rm \tau}]P_\theta[T^{\rm \tau}|T^{\rm \tau - 1:1}, a^{\rm \tau-1:1}, T_{\rm out}^{\rm \tau-1:1}]
\label{eq:lambda}
\end{eqnarray}
Note, the variables $T^{\rm t}, \dots, T^{\rm 1}$ are unobserved , meaning we have a so-called incomplete likelihood function. By incompleteness, we mean that the likelihood is the result of marginalization over some unobserved  variables $Z$: $P[X] = \int P[X, Z]dZ$ that we cannot estimate directly. We can use a well-known method based on evidence lower bound (ELBO) instead \cite{ELBO}. Shifting our goal to find the argmax of the incomplete likelihood function and the value of the unobserved  variables. 

In the ELBO approach, for any given incomplete likelihood, we can build the following lower bound:
\begin{eqnarray}
\log P[X] \geq \int dZ q[Z|X]\log\frac{P[X, Z]}{q[Z|X]}
\end{eqnarray}

\noindent where $q[Z|X]$ is any pdf over $Z$. The bound becomes exact if and only if $q[Z|X] = P[Z|X]$. In practice, one can use some parametric family to approximate $q[Z|X]$. The final formula reads:
\begin{eqnarray}
\log P_\theta[X]>\int dZ q_{\tilde{\theta}}[Z|X]\log\frac{P_\theta[X, Z]}{q_{\tilde{\theta}}[Z|X]}
\label{eq:parametric_ELBO}
\end{eqnarray}
where $\tilde{\theta}$ parametrizes the pdf over the unobserved  parameters. By maximizing ELBO over $\theta$ and $\tilde{\theta}$, we maximize our incomplete likelihood function and find the unobserved  variables. Now, going back to our case, the true indoor temperature $T$ plays the role of an unobserved variable that we aim to find. For convenience, we choose $q_{\tilde{\theta}}$ in the form:

\begin{equation}
q_{\tilde{\theta}}[T^{\rm t:1}| T_{\rm obs}^{\rm t:1}, a^{\rm t:1}, T_{\rm out}^{\rm t:1}] = \frac{1}{\sqrt{2\pi}\tilde{\sigma}_{\tilde{\theta}}^{\rm t}}\exp\left(-\frac{(T^{\rm t} - \tilde{\mu}_{\tilde{\theta}}^{\rm t})^2}{2\tilde{\sigma}_{\tilde{\theta}}^{\rm t 2}}\right)
\label{eq:q}
\end{equation}

\noindent where $\tilde{\mu}_{\tilde{\theta}}^{\rm  t}=\tilde{\mu}_{\tilde{\theta}}^{\rm t}(T_{\rm obs}^{\rm t:1}, T_{\rm out}^{\rm t:1}, a^{\rm t:1})$ and $\tilde{\sigma}_{\tilde{\theta}}^{\rm t}=\tilde{\sigma}_{\tilde{\theta}}^{\rm t}(T_{\rm obs}^{\rm t:1}, T_{\rm out}^{\rm t:1}, a^{\rm t:1})$ are also parametric families, which we choose to approximate in the form of a single one-dimensional convolution neural network (CNN) \cite{CNN}. For more details, see Section \ref{CNNdenoise}. $\tilde{\mu}_{\tilde{\theta}}^{\rm t}$ can be interpreted as our denoiser and $\tilde{\sigma}_{\tilde{\theta}}^{\rm t}$ as error of denoising. Similarly to Eq.~\eqref{eq:lambda}, to find the exact solution for the ELBO function we plug equations (\ref{eq:Pnoise}), (\ref{es:Ppred}) and (\ref{eq:q}) under the integral over time in Eq.~\eqref{eq:parametric_ELBO}. The ELBO function takes the following form:

\begin{eqnarray}
&&{\rm ELBO} = \int \prod_{\rm i=1}^t {\rm d}T^{\rm i} \underbrace{\frac{1}{(2\pi)^{t/2}\prod_{\tau=2}^t\tilde{\sigma}_{\tilde{\theta}}^{\rm \tau}}\exp\left(-\sum_{\tau=2}^t\frac{(T^{\rm \tau} - \tilde{\mu}_{\tilde{\theta}}^{\rm \tau})^2}{2\tilde{\sigma}_{\tilde{\theta}}^{\rm \tau 2}}\right)}_{q_{\tilde{\theta}}[T|T_{\rm obs}]\ {\rm from  \ Eq. \ (\ref{eq:q}) }}\nonumber \\
&&\times\Bigg[\log\Bigg\{ \prod_{\tau=2}^{t}\underbrace{\frac{1}{\sqrt{2\pi}\sigma}\exp\left(-\frac{(T_{\rm obs}^{\rm \tau}-T^{\rm \tau})^2}{2\sigma^2}\right)}_{P[T_{\rm obs}^{\rm \tau}|T^{\rm \tau}]{\rm \ from \ Eq. \ (\ref{eq:Pnoise})}}\underbrace{\frac{1}{\sqrt{2\pi}s}\exp\left(-\frac{(T^{\rm \tau} - \mu_\theta^{\rm \tau})^2}{2s^2}\right)}_{P_{{\theta}}[T^{\rm \tau}|T^{\rm \tau - 1:1}, a^{\rm \tau-1:1}, T_{\rm out}^{\rm \tau-1:1}] \ {\rm from \ Eq. \ (\ref{es:Ppred}) } }\Bigg\}\nonumber \\
&&-\log\Bigg\{\underbrace{\frac{1}{(2\pi)^{t/2}\prod_{\tau=2}^t\tilde{\sigma}_{\tilde{\theta}}^{\rm \tau}}\exp\left(-\sum_{\tau=2}^t\frac{(T^{\rm \tau} - \tilde{\mu}_{\tilde{\theta}}^{\rm \tau})^2}{2\tilde{\sigma}_{\tilde{\theta}}^{\rm \tau 2}}\right)}_{q_{\tilde{\theta}}[T|T_{\rm obs}] {\rm \ from  \ Eq. \ (\ref{eq:q})}} \Bigg\}\Bigg]
\label{eq:9}
\end{eqnarray}
This integral cannot be calculated analytically, but it can be estimated by sampling from $q$. To do so, we need to perform the following change of variables: $\epsilon^{\rm \tau} = (T^{\rm \tau} - \tilde{\mu}_{\tilde{\theta}}^{\rm \tau})/\tilde{\sigma}_{\tilde{\theta}}^{\rm \tau}$, and $T^{\rm \tau} = \epsilon^{\rm \tau}\tilde{\sigma}_{\tilde{\theta}}^{\rm \tau} + \tilde{\mu}_{\tilde{\theta}}^{\rm \tau}$. This change of variables is called the reparametrization trick \cite{VAE}. After the change of variables and calculating the integral, equation \ref{eq:9} can be expressed as:
{\small
\begin{eqnarray}
&&{\rm ELBO} = \Bigg\langle -\frac{t}{2}\log2\pi - t\log\sigma - t\log s + \sum_{\tau=2}^t\log \tilde{\sigma}_{\tilde{\theta}}^{\rm \tau} - \frac{1}{2\sigma^2}\sum_{\tau=2}^t\Big[T_{\rm obs}^{\rm \tau} - \epsilon^{\rm \tau}\tilde{\sigma}_{\tilde{\theta}}^{\rm \tau} - \tilde{\mu}_{\tilde{\theta}}^{\rm \tau}\Big]^2 \nonumber\\ && -\frac{1}{2s^2}\sum_{\tau=2}^t\bigg[\epsilon^{\rm \tau}\tilde{\sigma}_{\tilde{\theta}}^{\rm \tau} + \tilde{\mu}_{\tilde{\theta}}^{\rm \tau} - \mu_\theta^{\rm \tau-1}\bigg]^2 + \sum_{\tau=2}^t\frac{(\epsilon^{\rm \tau})^2}{2}\Bigg\rangle_{\epsilon\sim N(0, I)}
\label{eq:12}
\end{eqnarray}}
Finally, after partial averaging over $\epsilon$, equation (\ref{eq:12})  becomes the following expression:

\begin{eqnarray}
\nonumber
{\rm ELBO} &=& -\frac{t}{2}\log2\pi - t\log\sigma - t\log s + \sum_{\tau=2}^t\log \tilde{\sigma}_{\tilde{\theta}}^{\rm \tau}\\
&-& \frac{1}{2\sigma^2}\sum_{\tau=2}^t\Big[\big[T_{\rm obs}^{\rm \tau} - \tilde{\mu}_{\tilde{\theta}}^{\rm \tau}\big]^2 + (\tilde{\sigma}_{\tilde{\theta}}^{\rm \tau})^2\Big] \nonumber\\ & - &\frac{1}{2s^2}\sum_{\tau=2}^t\Bigg\{(\tilde{\sigma}_{\tilde{\theta}}^{\rm \tau})^2 + \Bigg\langle\bigg[\tilde{\mu}_{\tilde{\theta}}^{\rm \tau} - \mu_\theta^{\rm \tau - 1}\bigg]^2\Bigg\rangle_{\epsilon\sim N(0, I)}\Bigg\} + \frac{t}{2}
\end{eqnarray}

\noindent The average over $\epsilon$ left, can be estimated by sampling from the normal distribution with standard parameters. Then, using gradient ascent in the space of parameters $\theta$ and $\tilde{\theta}$, we find the optimal parameters of the   ${\rm ELBO}$ function for both parts: the denoiser and the predictor.  Therefore, the  ${\rm ELBO}$ function will play the role of our loss function. In practice, using the Variational Bayesian approach, both parts (dynamics predictor and denoiser) can be trained simultaneously (i.e. allowing error gradients to be backpropagated through both parts and optimizing a single, unified loss function). Our approach does not demand special probabilistic libraries, which can be computationally expensive, the sampling from the normal distribution for $\epsilon$, can be approximated from the extracted noise $\tilde{\sigma}_{\tilde{\theta}}^{\rm \tau}$.

\subsection{Neural network architecture}

The all-in-one method mentioned above includes two parts. The first part calculates the pdf of the denoised history of states with a given noisy history of states $P[s_{\rm denoised}^{\rm t}, \dots, s_{\rm denoised}^1|s_{\rm noisy}^{\rm t}, \dots, s_{\rm noisy}^1]$. This distribution plays the role of denoiser and filters out noise from historical data. We choose to parametrize this pdf using a one-dimensional convolutional neural network in the time domain, as practice shows  CNNs are exceptionally resilient to data distortion and computationally efficient \cite{KIRANYAZ2021107398}. Other neural network architectures, including RNNs and deep autoregressive networks, can also be considered as alternatives for parameterizing this pdf. The second part calculates the distribution over the next state of the system with a given history of previous states $P[s_{\rm denoised}^{\rm t+1}|s_{\rm denoised}^{\rm t}, \dots, s_{\rm denoised}^1]$. A RNN or a deep autoregressive network can parameterize this pdf, and the pdf plays the role of a predictor. Here, we choose a RNN, to parameterized this pdf as RNNs allow for controlling memory effects importance in the prediction process \cite{Haviv2019UnderstandingAC}, which is critical for modeling thermal inertia.  The whole architecture can be trained self-consistently by using the Bayesian approach. This architecture is the adaptation and generalization of the recently developed deep Kalman Filter for energy systems \cite{DKF} which shows excellent performance on the given problem. The particular implementation of this architecture and numerical experiments with this architecture are discussed extensively in the next section.

\subsubsection{Architecture of CNN for denoising}
\label{CNNdenoise}

\begin{figure}[ht]
    \centering
    \includegraphics[width=0.5\linewidth]{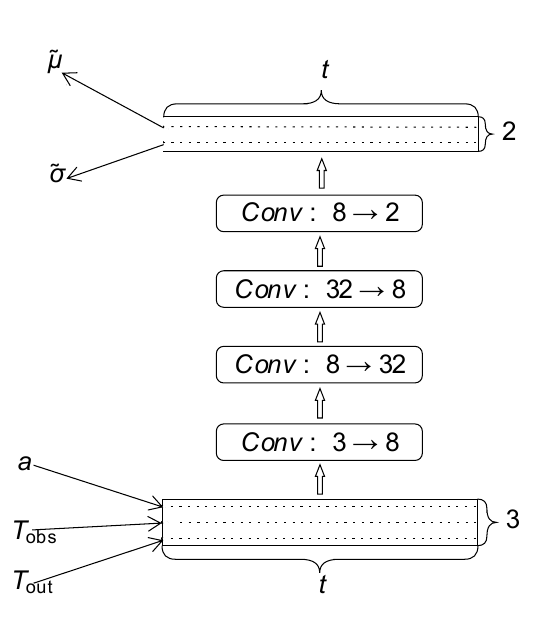}
    \caption{Architecture of the denoiser.}
    \label{fig3}
\end{figure}

We use a one-dimensional CNN as the denoiser. The denoiser takes as an input a concatenation of the actions array $a$, the observed temperature array $T_{\rm obs}$ and the outside temperature array $T_{\rm out}$. As a result, we get a matrix of shape $3\times t$. Then, this matrix goes through $4$ one-dimensional convolution layers. Finally, we get a denoised observed temperature $\tilde{\mu}$ and the standard deviation of observed noise $\tilde{\sigma}$. The diagrammatic representation of this chain of transformations is shown in Fig.~\ref{fig3}.

\subsubsection{Architecture of RNN for dynamics prediction}
\label{RNNarch}

\begin{figure}[ht]
    \centering
    \includegraphics{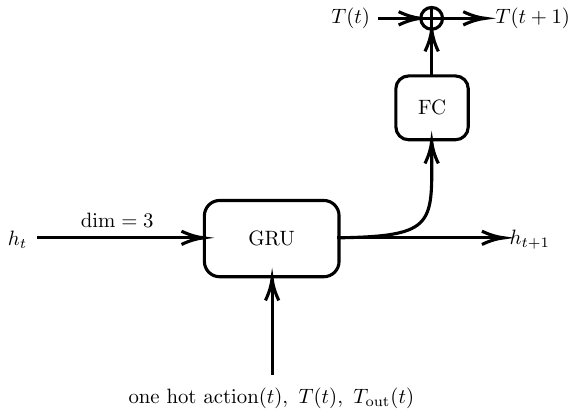}
    \caption{Architecture of one RNN cell}
    \label{fig1}
\end{figure}

\begin{figure}[ht]
    \centering
    \includegraphics{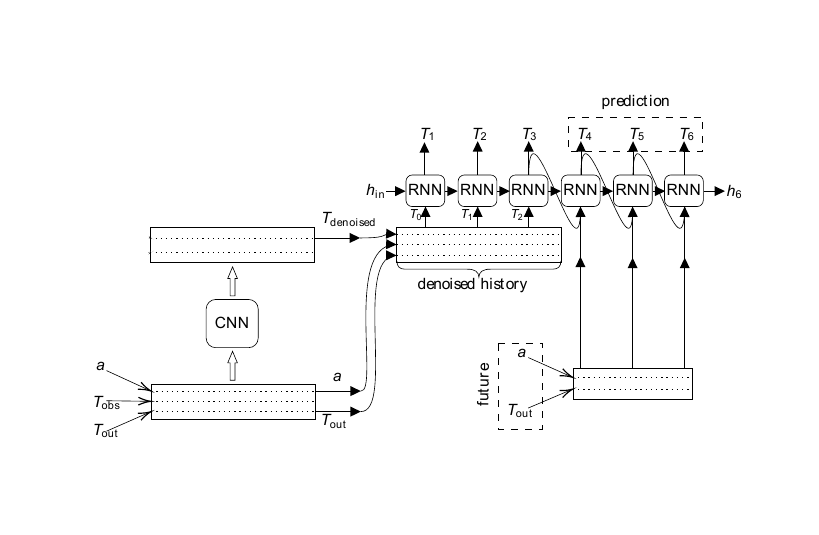}
    \caption{Schematic illustration of the data flow at the prediction stage.}
    \label{fig4}
\end{figure}

We approximate the function $\mu_\theta$, used as the parameters of the system dynamics using an RNN. The RNN cell in our scheme consists of a gated recurrent unit (GRU) \cite{GRU} and a small fully connected neural network. The dimension of the GRU hidden state is $3$. The GRU cell takes as an input a concatenation of the previous action vector codified in a one-hot vector of depth $4$, the previous temperature in the room and the previous outside temperature. Then the GRU cell's hidden state goes through a fully connected neural network and returns an update of the previous temperature. Finally, this update and the previous temperature are summed up thus providing the prediction of the next step temperature. A graphical representation of this architecture is shown in Fig.~\ref{fig1}. The whole data flow, including the chain of RNN cells, is illustrated in Fig.~\ref{fig4}.


\subsubsection{Training data}
The modelled environment is dynamic and includes realistic low-noise simulations of a wide range of outdoor environments. The dataset includes a range of noise levels from 0.01 up to 0.25 standard deviations, the upper bound was selected as it represents a maximum cumulative error of one degree Celsius, corresponding to the noise level of the cheapest among popular Arduino sensors as the DHT11 or TMP36. The different noise levels help the deep Kalman filter to generalize observed data from many possible sources. This feature allows our model to be deployed in different setups without losing performance. The dataset was created using different starting conditions for the simulations with a length of 3000 (1500) minutes, later each simulation was divided into 10 (5) sequences of 300 minutes. This approach has the advantage that only a fraction of the starting points of the simulations are determined by the user, while most starting points are chosen randomly from nearby values, therefore the user cannot introduce bias into the simulations by mistake. Also, it allows the simulation of separate moments of the day-night cycle. All the defined model parameters in section \ref{simulation}, which constitute the state space of our problem were bounded and uniformly sampled as shown in Table \ref{tab:dataset}. As a final step a subset was randomly selected for the subsequent simulations. Similarly, a 2400-minute validation dataset, a fine-tuning dataset, a demand response dataset and a fault dataset were created with 500  simulations each.

After training, we can use our model to predict the dynamics of the indoor temperature. The prediction consists of three parts. First, we pass history through the denoiser to extract true indoor temperature. Then we pass cleaned history through RNN to reconstruct the hidden state of RNN before prediction. Finally, we predict dynamics using a preliminary calculated hidden state, the future sequence of control signals, and the prediction of outdoor temperature.

\begin{table*}[t]
    \centering
    \begin{tabular}{|c|ccc|ccc|}
    \hline
 Parameter& \multicolumn{3}{c|}{Training dataset}& \multicolumn{3}{c|}{Validation dataset}\\
 \cline{2-7}
 & Start & End & Samples &  Start & End & Samples\\
        \hline
        \hline
        Noise level [p.u.]& 0.01 & 0.25 & 5 & 0.01 &  0.25  & 9\\
        Indoor temperature  [K]&  293.15&  303.15  & 5 & 293.15  & 303.15  & 8\\
        Ventilation level [-] & 20  &  60 & 4 & 20 &  60& 10\\
        Heater temperature [K]& 293.15  & 340.15  & 5 & 293.15  & 340.15  & 8\\
        Wall temperature [K]& 273.15  & 308.15  & 10 & 273.15  & 308.15 &  14\\
        Outside temperature [K]& 253.15   & 308.15 & 30 & 253.15 &  308.15  & 39\\
        \hline
        \hline
        Total size & \multicolumn{3}{c|}{170000}& \multicolumn{3}{c|}{3467520}\\
        Sequences per simulation & \multicolumn{3}{c|}{10}& \multicolumn{3}{c|}{5}\\
        Selected subset& \multicolumn{3}{c|}{10000 (100000)}& \multicolumn{3}{c|}{9000 (45000)}\\
        \hline
    \end{tabular}
    \caption{Range, sampling density and size of the training and validation datasets.}
    \label{tab:dataset}
\end{table*}

\section{Results}
\label{sec:Results}

In this section, we discuss the results achieved by training the proposed neural network, including  synthetic dataset description, performance metrics, fine-tuning and specialization strategies, robustness to measurement noise, and finally two application cases outside typical predictive control: thermal DR event length prediction and fault detection in HVAC and thermal envelope.

\subsection{Indoor temperature prediction}
\subsubsection{Simulation validation}
\begin{figure*}[t!]
    \centering
    \begin{subfigure}[t]{0.46\textwidth}
        \centering
        \includegraphics[width=\linewidth]{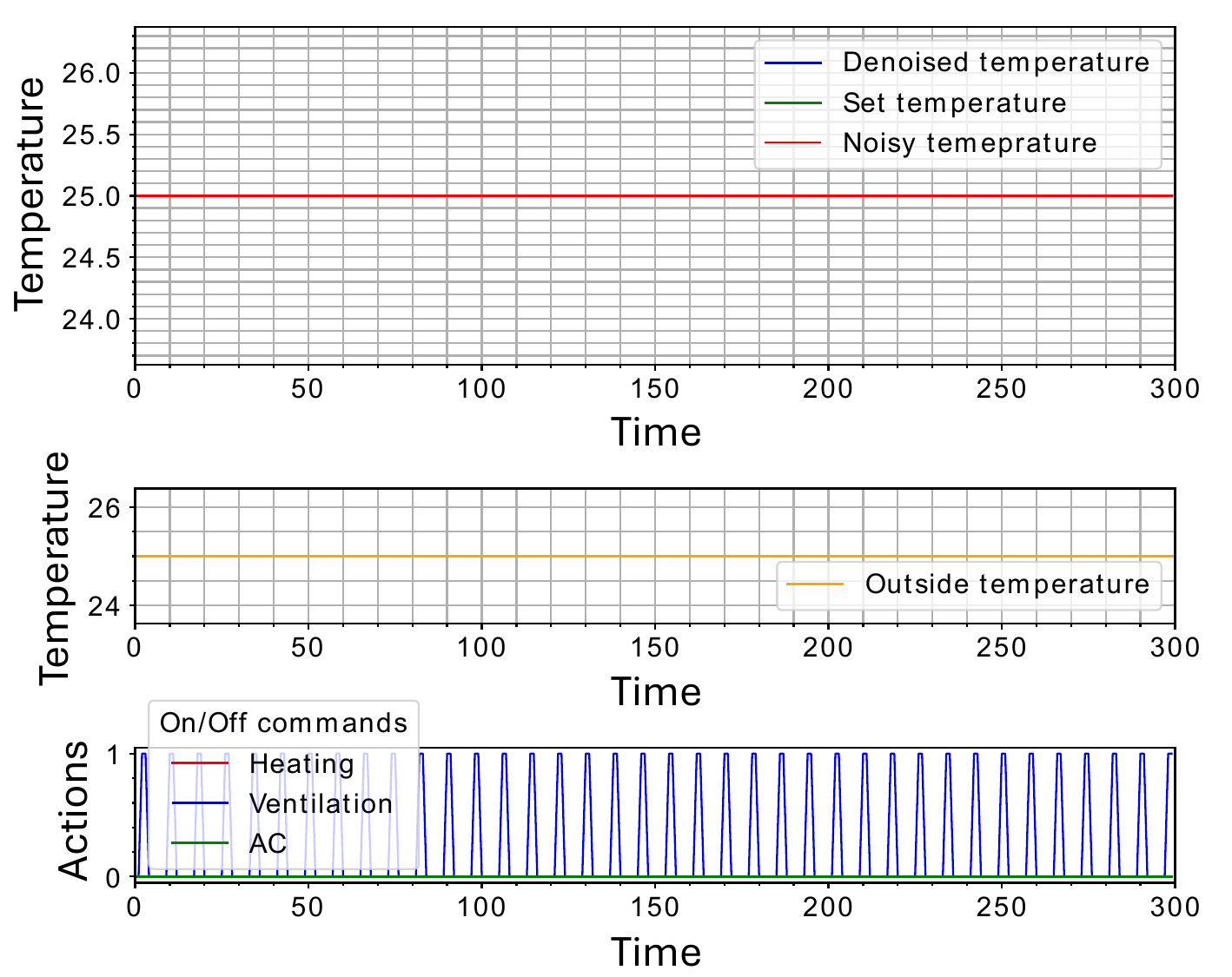}
        \caption{Scenario 1}
    \end{subfigure}%
    \hspace{0.01\textwidth} 
    \begin{subfigure}[t]{0.46\textwidth}
        \centering
        \includegraphics[width=\linewidth]{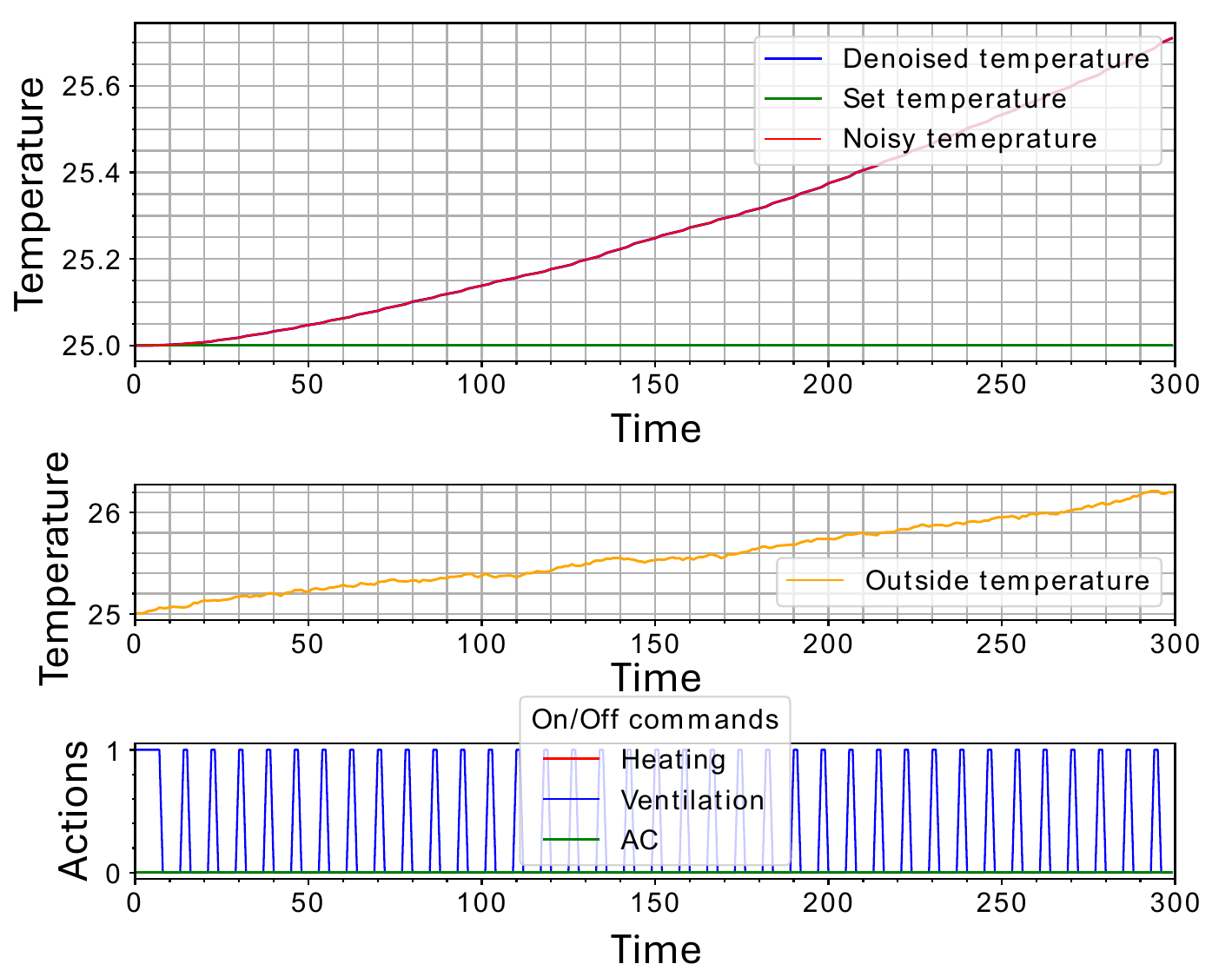}
        \caption{Scenario 2}
    \end{subfigure}
    
    \begin{subfigure}[t]{0.46\textwidth}
        \centering
        \includegraphics[width=\linewidth]{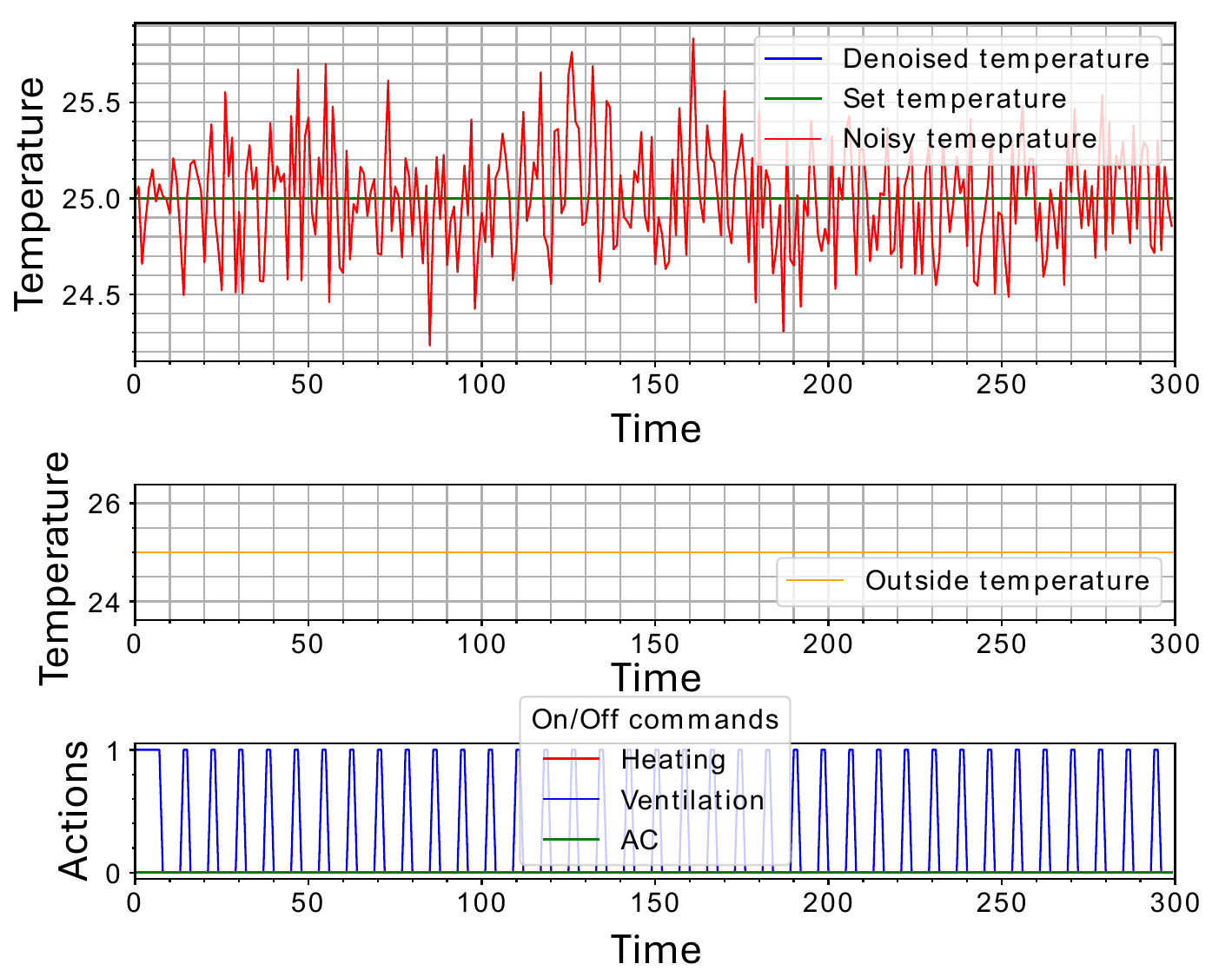}
        \caption{Scenario 3}
    \end{subfigure}%
    \hspace{0.01\textwidth} 
    \begin{subfigure}[t]{0.46\textwidth}
        \centering
        \includegraphics[width=\linewidth]{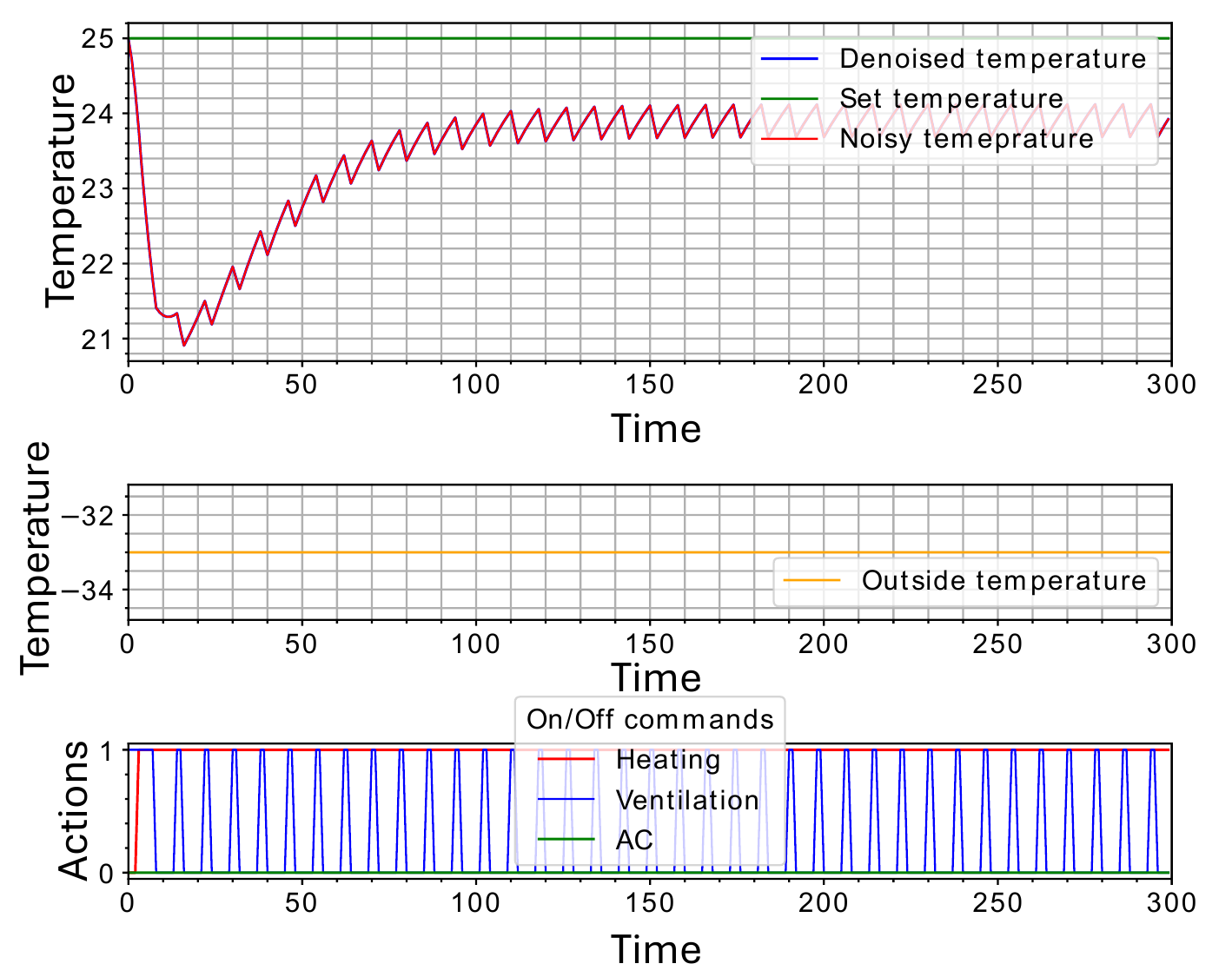}
        \caption{Scenario 4}
    \end{subfigure}
    
    \begin{subfigure}[t]{0.46\textwidth}
        \centering
        \includegraphics[width=\linewidth]{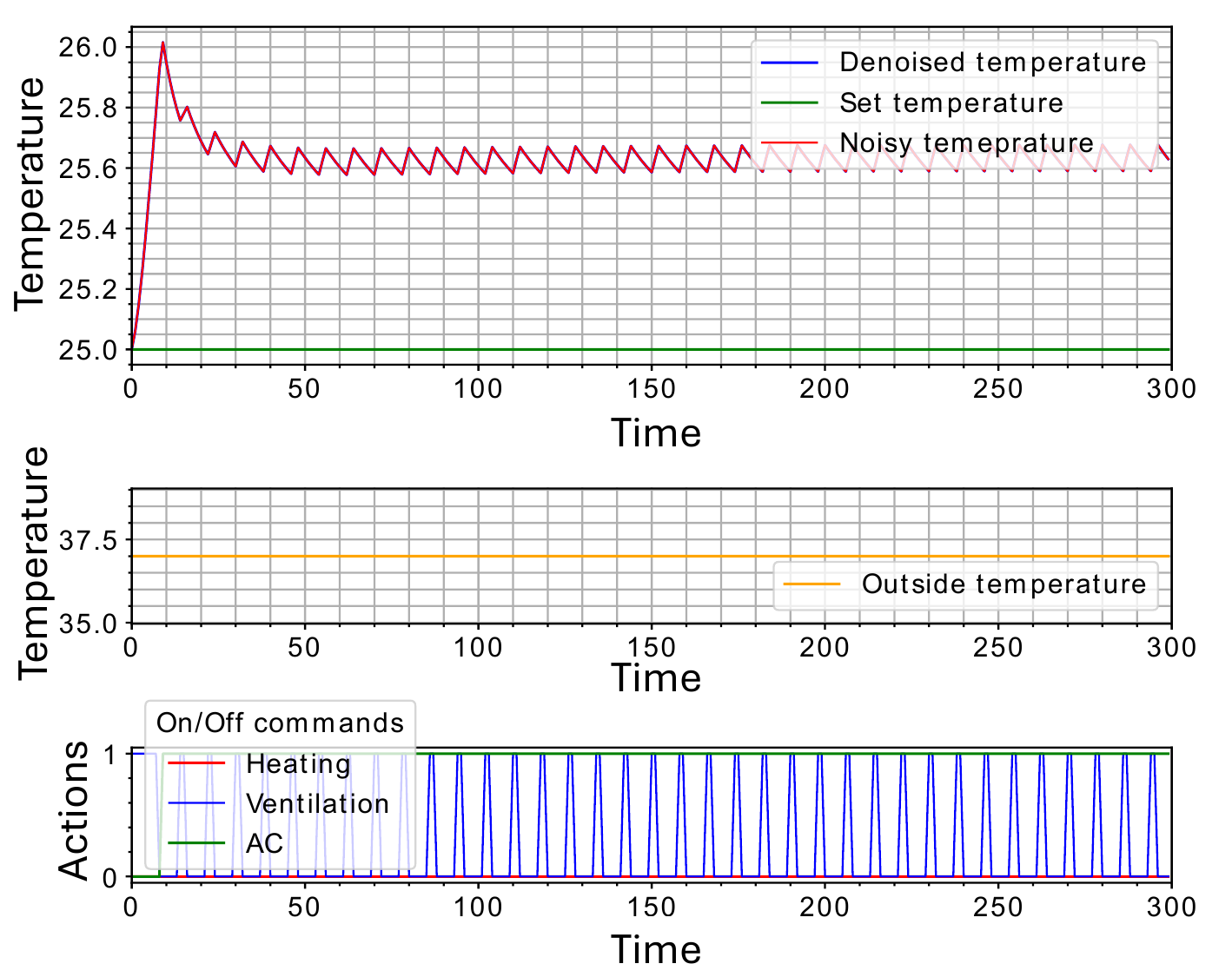}
        \caption{Scenario 5}
    \end{subfigure}%
    \hspace{0.01\textwidth} 
    \begin{subfigure}[t]{0.46\textwidth}
        \centering
        \includegraphics[width=\linewidth]{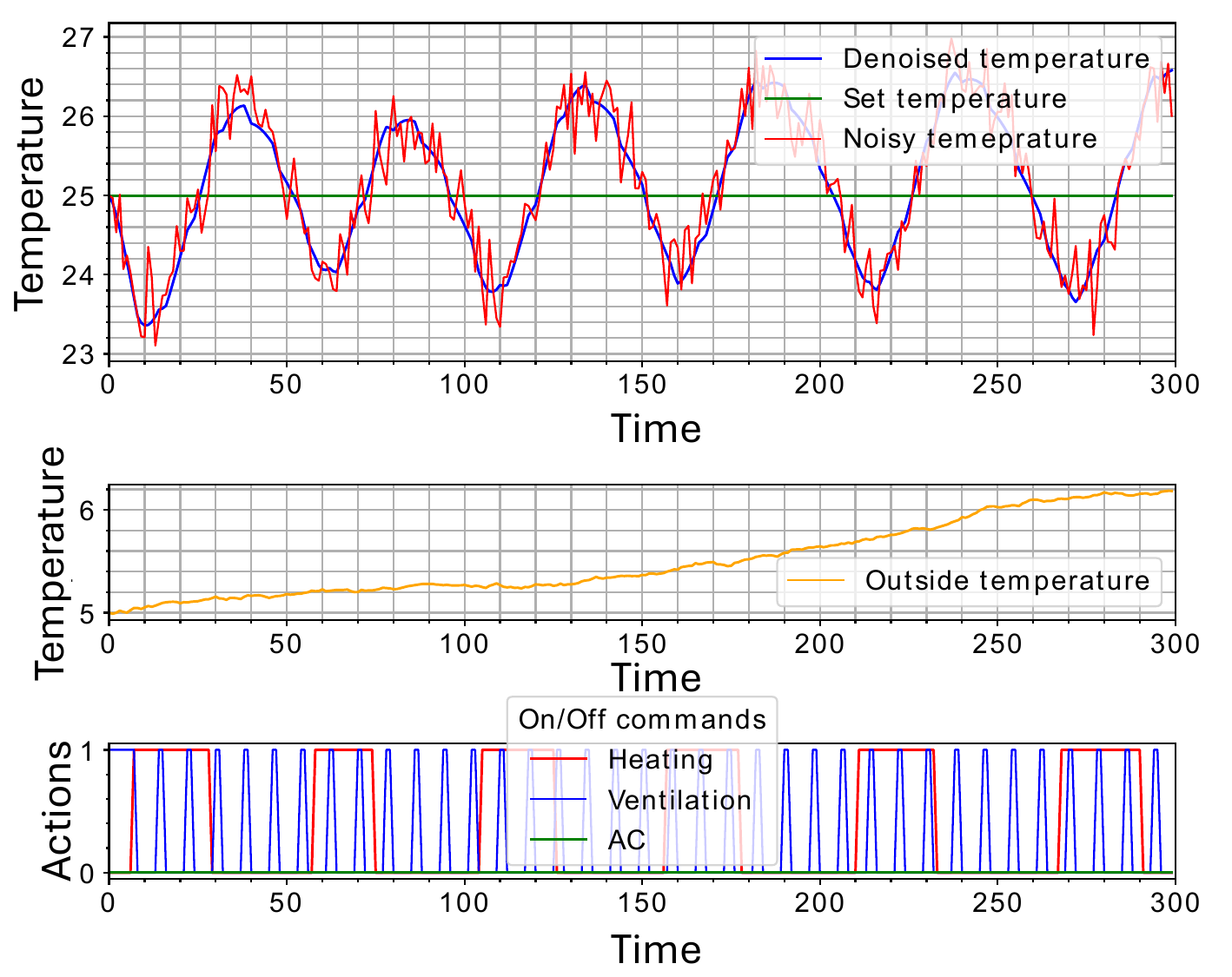}
        \caption{Scenario 6}
    \end{subfigure}
    
    \caption{Simulation results for different scenarios.}
    \label{fig:simulation_results}
\end{figure*}

Our model simulates a single-room building in a ``warm summer humid continental climate'' capable of maintaining a set temperature of 25 $^{\circ}$C with an outside temperature range from $-$30$^{\circ}$ C to 35$^{\circ}$ C. To validate the correct behaviour of our model we look for expected behaviour using the scenarios described in Table \ref{tab:scenarios}. Graphical demonstrations of correct behaviour are shown in Fig.~\ref{fig:simulation_results}. Note that we use simulated minute weather data. Note that such a forecast data is rare in current applications; however, it is becoming increasingly available due to the benefits it represents for solar and wind power generation \cite{WANG2020113075,JUNCKLAUSMARTINS2022100019}. Although 5-, 10- and 15-minute forecast data are widely accessible and can be used with our model, this would take a toll on its performance.

\begin{table}[ht!]
\centering
\begin{tabular}{|P{1.3cm}|c|P{1.3cm}|P{4cm}|P{4cm}|}
\hline
\textbf{Scena-rio} & \textbf{Noise} & \textbf{$T_{\rm out}=$ constant} &\textbf{Temperature \hspace{1cm} relation} & \textbf{Expected outcome} \\ \hline
1 & No & Yes & \( T_{\rm out} = T_{\rm set} = T_{\rm obs} \) & \( T_{\rm obs} = T_{\rm set} =\) constant \\ \hline
2 & No & No & \( T_{\rm out} \approx T_{\rm set} \approx T_{\rm obs} \) & \( T_{\rm obs} \) follows \( T_{\rm out} \). \\ \hline
3 & Yes & Yes & \( T_{\rm out} = T_{{set}} = T_{\rm obs} \) & \( T_{\rm obs} \approx T_{\rm set} \) (with noise). \\ \hline
4 & No & Yes & \( T_{\rm out} \ll T_{\rm set} \approx -30^\circ\)C & \( T_{\rm obs} < T_{\rm set} \) (cannot reach set point). \\ \hline
5 & No & Yes & \( T_{\rm out} \gg T_{\rm set} \approx 35^\circ \)C & \( T_{\rm obs} > T_{\rm set} \) (cannot reach set point). \\ \hline
6 & Yes & Yes & \( T_{\rm out} = T_{\rm set} - 20^\circ \)C & Noise introduces variations in \(T_{\rm obs} \); the control signal is stochastic. \\ \hline
\end{tabular}
\caption{Scenarios with expected outcomes under varying conditions. Here, $T_{\rm set}$ refers to the target temperature for the control system.}
\label{tab:scenarios}
\end{table}

\subsubsection{Performance metrics}
Standard metrics for time series prediction are RMSE, MAE and the coefficient of determination ($R^2$). The presented metrics are used to evaluate our model's prediction performance at different time horizons: 30, 60, 90, 120, 150 and 2400 minutes. The results are presented in Table \ref{tab:typical_metrics}. We can observe that the quality of the model is overall consistent for different prediction horizons, and allows for a wide implementation of tasks that require predictions from tens of minutes to several hours ahead. To evaluate the performance of our model, we compared it with state-of-the-art models across different metrics. Regarding MSE metric, several models compared by Afroz et al.~\cite{Real_time_prediction2018}, report values ranging from 0.003 to 0.18, depending on the prediction horizon  (for short 5-minute predictions up to 3-hour predictions). In comparison, our model achieves MSE values varying from 0.038 to 0.0762. Among the models compared in \cite{Real_time_prediction2018}, only the one proposed by Afroz et al., outperforms our model in the few-hours prediction task. As regards the RMSE, state-of-the-art models typically report values around 0.6$^{\circ}$ C \cite{delcroix2021autoregressive,fang2021multi}, while the recent model proposed by Bampoulas et al. boasts an outstanding RMSE of 0.188$^{\circ}$ C for an hour-ahead prediction and 0.42$^{\circ}$ C for a day-ahead prediction \cite{BAMPOULAS2023121576}. Since our model achieves a RMSE ranging from 0.197$^{\circ}$ C for 30 min prediction horizon to 0.276$^{\circ}$ C for a 2400 min prediction horizon, its performance is comparable if not better than most available models. Note that our model performance remains stable for different prediction horizons compared to all the reviewed models. Further, it must be noted that, to the best of our knowledge, a benchmarking framework for such models is still lacking. Indeed, full performance comparison with works such as, e.g., that of Lopez-Villamor et al., who predicts HVAC load for commercial buildings with an $R^2$ value of 0.965 \cite{LOPEZVILLAMOR2024100777}, or that of Shi et al., who propose a deep RL algorithm for accounting for individual thermal comfort preferences achieving energy savings up to 39\% \cite{SHI2024118995}, cannot be directly made in spite of similarities among the works.  We present validation tests of our model using open source datasets in section \ref{sec:validation}.

\begin{table}[t]
    \centering
    \begin{tabular}{|c|ccc|}
    \hline
        Prediction horizon & RMSE& MAE& R$^2$\\
        \hline\hline
         30 &  0.1968& 0.128& 0.714\\
         60 &  0.2096& 0.137& 0.833\\
         90 &  0.2248& 0.147& 0.919\\
         120 & 0.2365& 0.155& 0.922\\
         150 & 0.2455& 0.162& 0.926\\
         500 & 0.2640 & 0.196 & 0.9139 \\
        1000 & 0.2708 & 0.203 & 0.908 \\
        1500 & 0.2741 & 0.207 & 0.905 \\
        2000 & 0.2760 & 0.209 & 0.904 \\
        2400 & 0.2762& 0.209 & 0.904 \\
        \hline\hline

\end{tabular}
\caption{Value of typical evaluation metrics for time series prediction scoring for different prediction horizons.}
    \label{tab:typical_metrics}
\end{table}

 Another way to measure the quality of the model is a fidelity band, where we can show the percentage of the validation dataset in which the prediction deviates from the true data less than a threshold. We measured for thresholds of $\pm$ 1$^{\circ}$ C, 0.5$^{\circ}$ C and 0.25$^{\circ}$ C. We include a parameter $n$ for the amount of consecutive out-of-range measurements that we think of as negligible, this method is suitable for tasks aiming to detect when the prediction diverges from the fidelity band, not to detect when it steps out of the band. We used $n=$ 0, 5, 10, 15, 20, and 30. The resulting share of the validation dataset in the  fidelity band range for any given prediction horizon and $n$, is given in Table \ref{tab:fidelity}. The data reveals that our model is suitable for predictions with a fidelity up to $\pm$ 0.5$^{\circ}$ C. As a rule of thumb the model consistently performed well in a fidelity band $\leq$ 2 RMSE. 

\begin{table*}[t]
\tiny
    \centering
      \begin{tabular}{|c|c|c|c|c|c|c|c|c|c|c|c|}
      \hline
 $n$&Fidelity & \multicolumn{10}{c|}{             Prediction horizon}\\
    \cline{3-12}
  &  band ($^\circ$C)  & 30     & 60     & 90     & 120    & 150    & 500    & 1000   & 1500   & 2000  & 2400   \\
\hline
   & 1                           & 0.991& 0.983& 0.980& 0.978& 0.975& 0.966& 0.94& 0.919& 0.904& 0.896\\
0  & 0.5                         & 0.838& 0.750& 0.682& 0.628& 0.594& 0.518& 0.478& 0.442& 0.436& 0.435\\
   & 0.25                        & 0.469& 0.339& 0.304& 0.281& 0.261& 0.1& 0.041& 0.022& 0.013& 0.012\\
   \hline
   & 1                           & 0.996& 0.991& 0.988& 0.985& 0.983& 0.996& 0.991& 0.984& 0.981& 0.979\\
5  & 0.5                         & 0.955& 0.907& 0.867& 0.826& 0.792& 0.596& 0.554& 0.504& 0.494& 0.492\\
   & 0.25                        & 0.673& 0.497& 0.44& 0.401& 0.376& 0.214& 0.139& 0.066& 0.057& 0.051\\
   \hline
   & 1                           & 1.000& 0.996& 0.993& 0.989& 0.988& 0.999& 0.999& 0.998& 0.998& 0.998\\
10 & 0.5                         & 0.986& 0.965& 0.947& 0.924& 0.904& 0.729& 0.661& 0.612& 0.593& 0.586\\
   & 0.25                        & 0.9& 0.753& 0.651& 0.560& 0.512& 0.28& 0.211& 0.139& 0.116& 0.109\\
   \hline
   & 1                           & 1.000& 0.998& 0.996& 0.996& 0.995& 1& 1& 0.999& 0.999& 0.999\\
15 & 0.5                         & 0.996& 0.989& 0.984& 0.979& 0.973& 0.889& 0.842& 0.798& 0.768& 0.758\\
   & 0.25                        & 0.968& 0.901& 0.834& 0.757& 0.702& 0.369& 0.262& 0.195& 0.178& 0.173\\
   \hline
   & 1                           & 1.000& 0.999& 0.996& 0.996& 0.996& 1& 1& 0.999& 0.999& 0.999\\
20 & 0.5                         & 0.997& 0.990& 0.985& 0.980& 0.975& 0.895& 0.848& 0.808& 0.775& 0.767\\
   & 0.25                        & 0.977& 0.915& 0.855& 0.787& 0.735& 0.434& 0.324& 0.249& 0.219& 0.209\\
   \hline
   & 1                           & 1.000& 0.999& 0.997& 0.996& 0.996& 1& 1& 0.999& 0.999& 0.999\\
25 & 0.5                         & 0.999& 0.995& 0.991& 0.986& 0.984& 0.972& 0.952& 0.932& 0.908& 0.902\\
   & 0.25                        & 0.998& 0.976& 0.951& 0.921& 0.883& 0.556& 0.419& 0.317& 0.284& 0.273\\
   \hline
   & 1                           & 1.000& 1.000& 0.998& 0.997& 0.997& 1& 1& 1& 1& 1\\
30 & 0.5                         & 1.000& 0.998& 0.995& 0.991& 0.988& 0.99& 0.979& 0.966& 0.954& 0.948\\
   & 0.25                        & 1& 0.991& 0.983& 0.968& 0.955& 0.732& 0.585& 0.477& 0.433& 0.413\\
        \hline
    \end{tabular}
    \caption{Share of the validation set, where the prediction falls inside the fidelity band of $\pm$ 1,0.5 and 0.25 K, for the prediction horizons 30,60,90,120,150,500,1000,1500,2000 and 2400 minutes, and for the different $n$  consecutive measurements up to 0,5,10,15,20,25 and 30 minutes considered negligible. }
    \label{tab:fidelity}
\end{table*}
\subsubsection{Measurement noise }

\begin{figure}[ht]
	\centering
	\includegraphics[width=\textwidth]{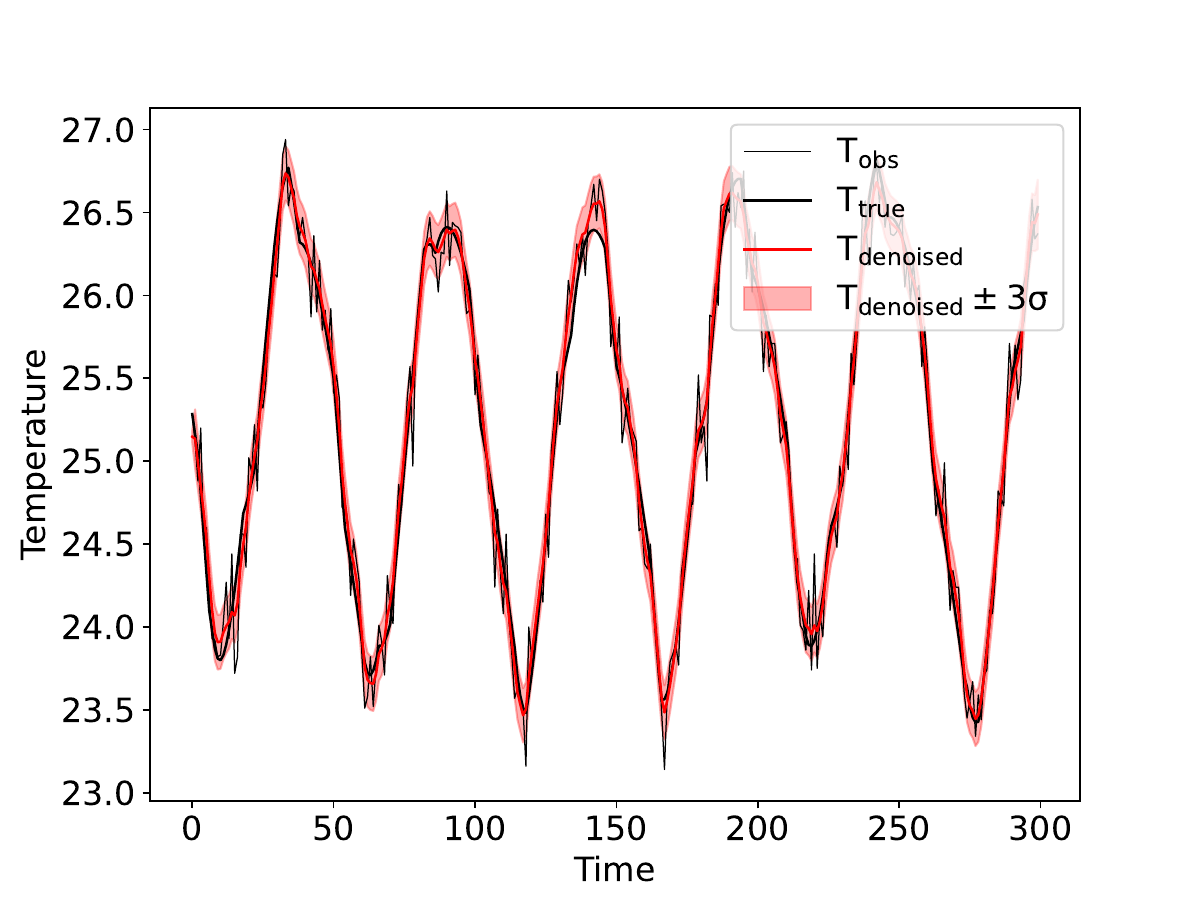}
	\caption{Example of denoised data using the deep Kalman filter without prediction. With true and observed temperatures in black. The predicted temperature is in red. The red zone corresponds to the trust region of the prediction.}

	\label{fig:Filtering}
\end{figure}

The deep Kalman filter trained in a wide range of noise levels ensures that our model is robust against measurement noise. This feature makes the model widely applicable on the condition that the predominant noise follows a normal distribution, hence allowing it to be deployed independently of the sensor quality, whether it is an indoor temperature sensor or an analogical sensor if used in other settings. The control system observes the noise levels influencing the behaviour of the HVAC system, so for our model to be resilient to measurement noise, means to have a high-performance adaptable Kalman filter and also not to rely too much on indoor temperature values. This is important since a shift of trends (i.e., a change of state on the heating or cooling equipment) may occur at different true temperatures $\rm T$. As shown in Table~\ref{tab:noise}, the quality of our model does not depend on the measurement noise levels within the range of noise used in the training data. The model is highly robust to measurement noise, that it can perform well with test data with unseen noise levels during the training stage  (200\% larger variance than during training ). However, the capability to reproduce the exact shape of the indoor temperature, including weak effects detection such as the cooling (heating) effect of ventilation, becomes compromised with noise levels outside the training dataset boundaries. An example of the denoising capabilities of the deep Kalman filter is shown in Fig.~\ref{fig:Filtering}.

\begin{table}
    \centering
    \begin{tabular}{|c|ccc|}
    \hline
         Noise&  RMSE&  MAE& $R^2$\\
         \hline
         0.01&   0.037&  0.0215& 0.998\\
         0.1&  0.057&   0.043&  0.996\\
         0.3&    0.136&  0.108& 0.978\\
 0.6& 0.264& 0.211&0.876\\
 0.8& 0.348& 0.278&0.744\\
 \hline
    \end{tabular}
    \caption{Model performance for different noise values. Noise is measured in standard deviations.}
    \label{tab:noise}
\end{table}

\subsubsection{Fine-tuning and specialization strategies}
\begin{figure}
    \centering

\begin{subfigure}{.49\textwidth}
    \includegraphics[width=\linewidth]{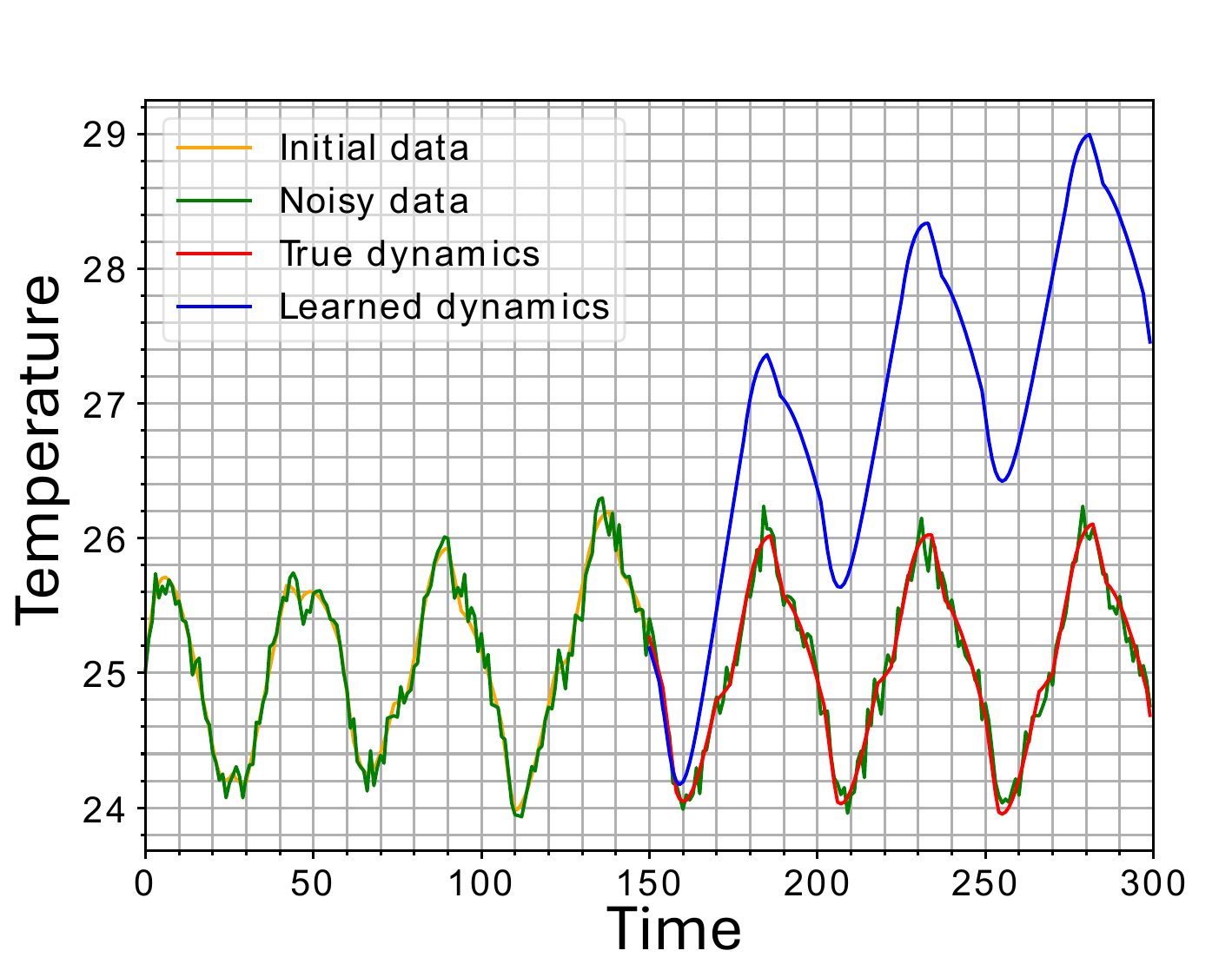}
    \caption{MAE: 2.03  RMSE: 2.19 R$^2$: -11.13  }
    \label{fig:finetuning_bf}
\end{subfigure}
\begin{subfigure}{.49\textwidth}
    \includegraphics[width=\linewidth]{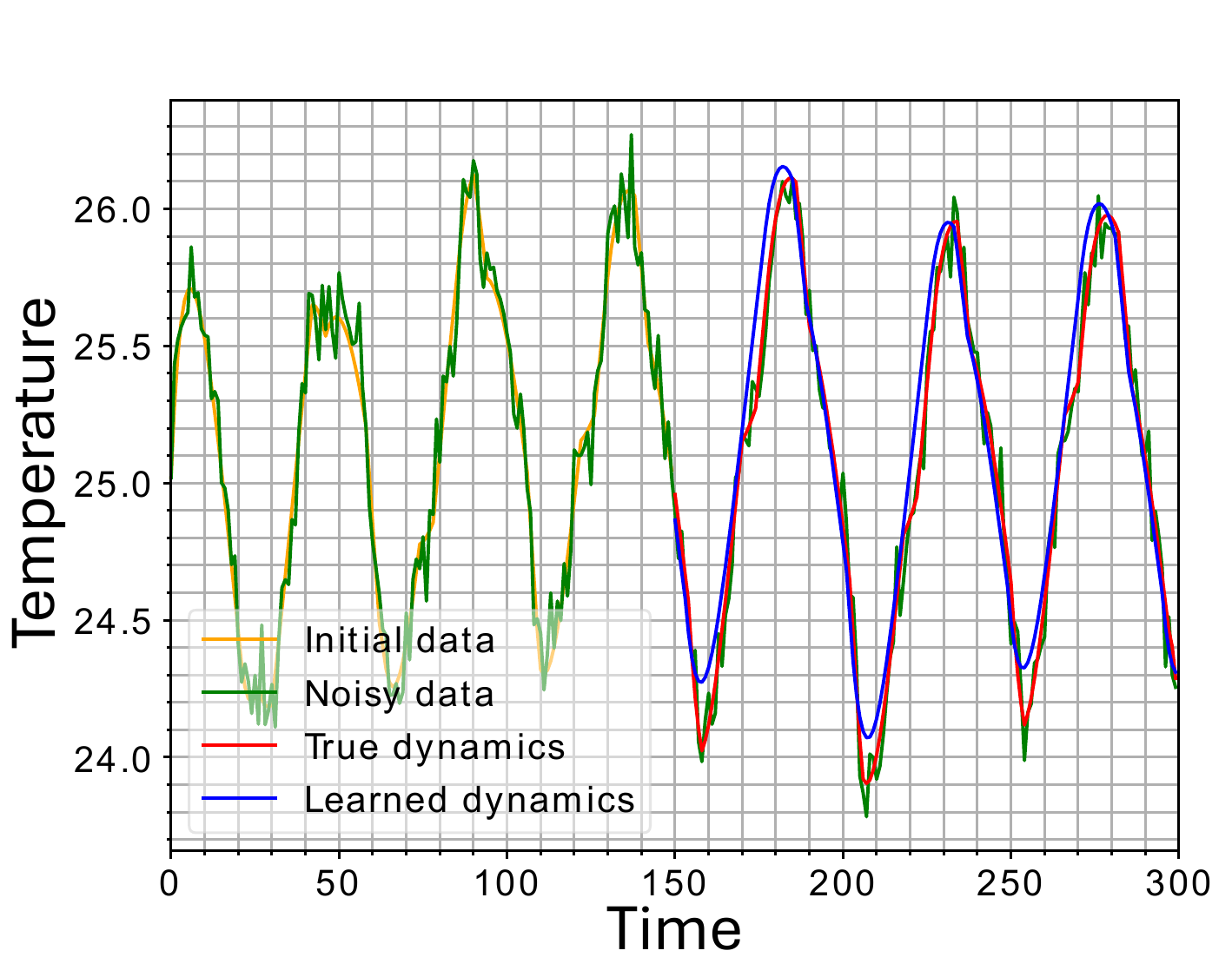}
    \caption{MAE: 0.1  RMSE: 0.12  R$^2$: 0.96 }
    \label{fig:finetuning_af}
\end{subfigure}
    \caption{Example of prediction using the model with and without fine-tuning. On the left panel, the simulated building has a single different parameter with a 15\% change from the parameters used for training. The right panel depicts the fine-tuned model on a two years dataset.}
    \label{fig:finetuning}
\end{figure}

As is the case for most ML models, fine-tuning is an important step in the model's exploitation. Tweaking our simulation in a single parameter within a 15\% change can illustrate how the model performance falls significantly even when deployed in a very similar environment, this is shown in figure \ref{fig:finetuning}. In a 500 samples test, the model metrics declined significantly from MAE: 0.162 RMSE: 0.2455 R$^2$: 0.926 as shown in Table \ref{tab:typical_metrics} to MAE: 1.35 RMSE: 1.97 R$^2$: $-3.68$.

During the fine-tuning stage, we generated a new training dataset with the same parameters shown in Table \ref{tab:dataset}, but with a single value for noise level standard deviation fixed at 0.25, i.e. simulating a sensor with a large fixed error rate and a low degradation rate. The dataset generated is equivalent to 1.5 months of data collection, including all the seasons of the year (i.e. a dataset that samples a whole day once per week), so it resembles an evenly distributed sample of the year-round dataset. Smaller datasets could be used to fine-tune the model, as long as it represents the full indoor-outdoor dynamics variability. Note that is recommended to divide the data into small chunks and shuffle it, so no concrete seasonal dynamics become dominant in the model. After 4000 iterations with a learning rate of 0.0001, we tested the model in a 500-sample dataset the metrics correspond to the quality of the original precision of the model on the original dataset MAE: 0.16 RMSE: 0.204  R$^2$: 0.851. The fine-tuned model may fail to reproduce the fine details such as the effects of ventilation if they are significantly different from those in the initial training dataset.

Another option to fine-tune the model is to use one fine-tuned model for cooling and another for heating, or even a third model for the dead band. Integrating them into production can be done with a simple if/else logic to select the correct model. This option has several advantages, including fast-tracking the model implementation via partial implementation of the model, which can be useful for cases with a lack of historical data. Another advantage is the increased precision of the models as they would record only the dynamics of specific equipment, such as thermal inertia which tend to be higher in heating systems than in cooling systems. For a single model learning both behaviours comes with a reduced prediction quality.

We exemplify this strategy with a heating-only and a cooling-only model using the same fine-tuning strategy. For the cooling-only fine-tuned model we could achieve  MAE: 0.13 RMSE: 0.164 R$^2$: 0.893, and  for  heating-only   MAE: 0.20 RMSE: 0.27 R$^2$: 0.92.

\subsection{Applications}
\label{sec:applications}

One very useful application of our model is to calculate the amount of time we could disconnect the HVAC system to participate in a demand response (DR) program. Another application is failure detection in the HVAC equipment, as the HVAC actions are input to our model, when the real data diverges from the fine-tuned model, we can detect a change in the behaviour of the HVAC and alert the maintenance personnel. Given that greenhouses are a few hours away from major cities, early fault detection can greatly reduce production losses.

\subsubsection{DR in thermal loads}
\label{sec:DR}

\begin{figure*}
\centering
\begin{subfigure}{.5\textwidth}
  \centering
  \includegraphics[width=\linewidth]{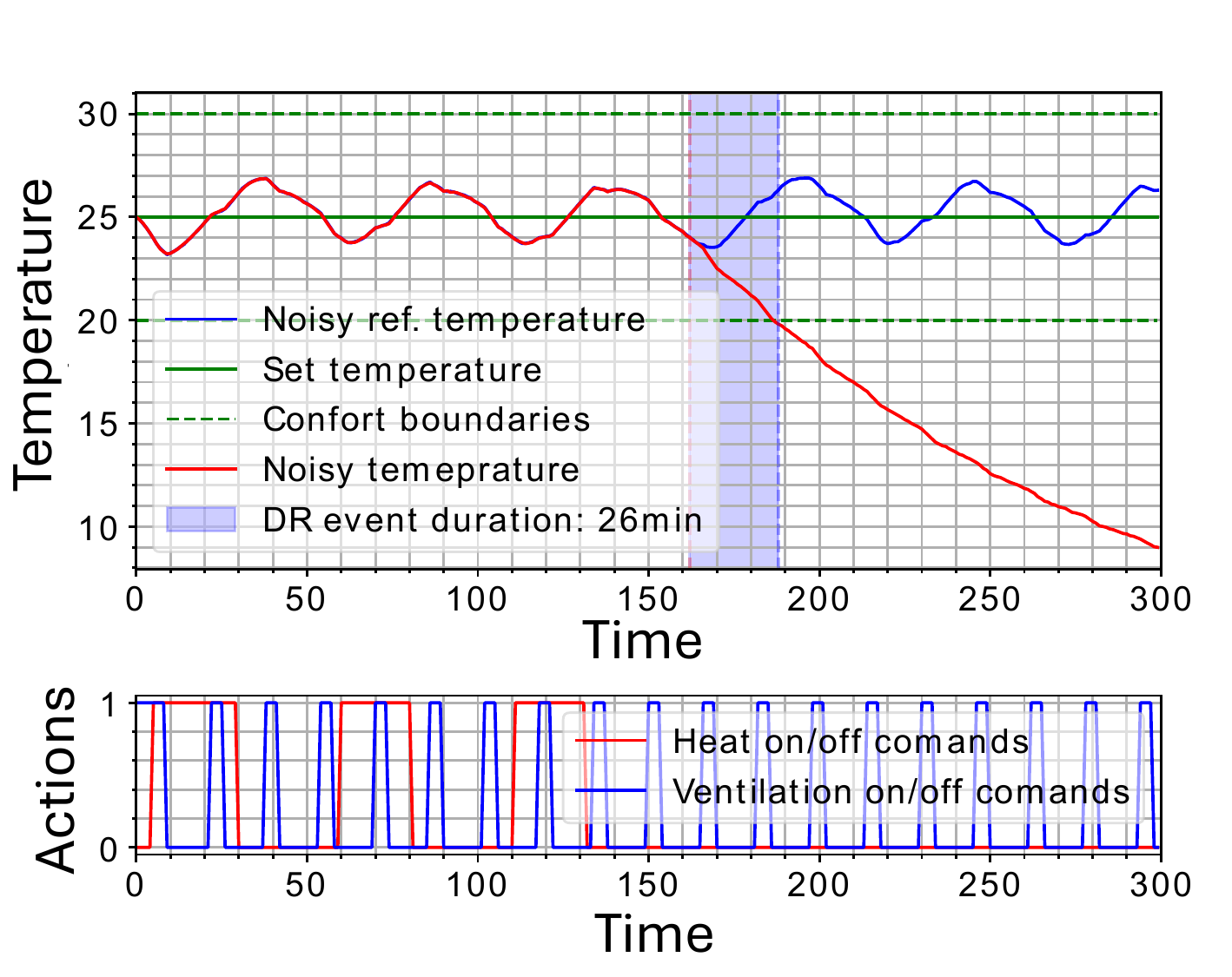}
  \caption{}
  \label{fig:sub1}
\end{subfigure}%
\begin{subfigure}{.5\textwidth}
  \centering
  \includegraphics[width=\linewidth]{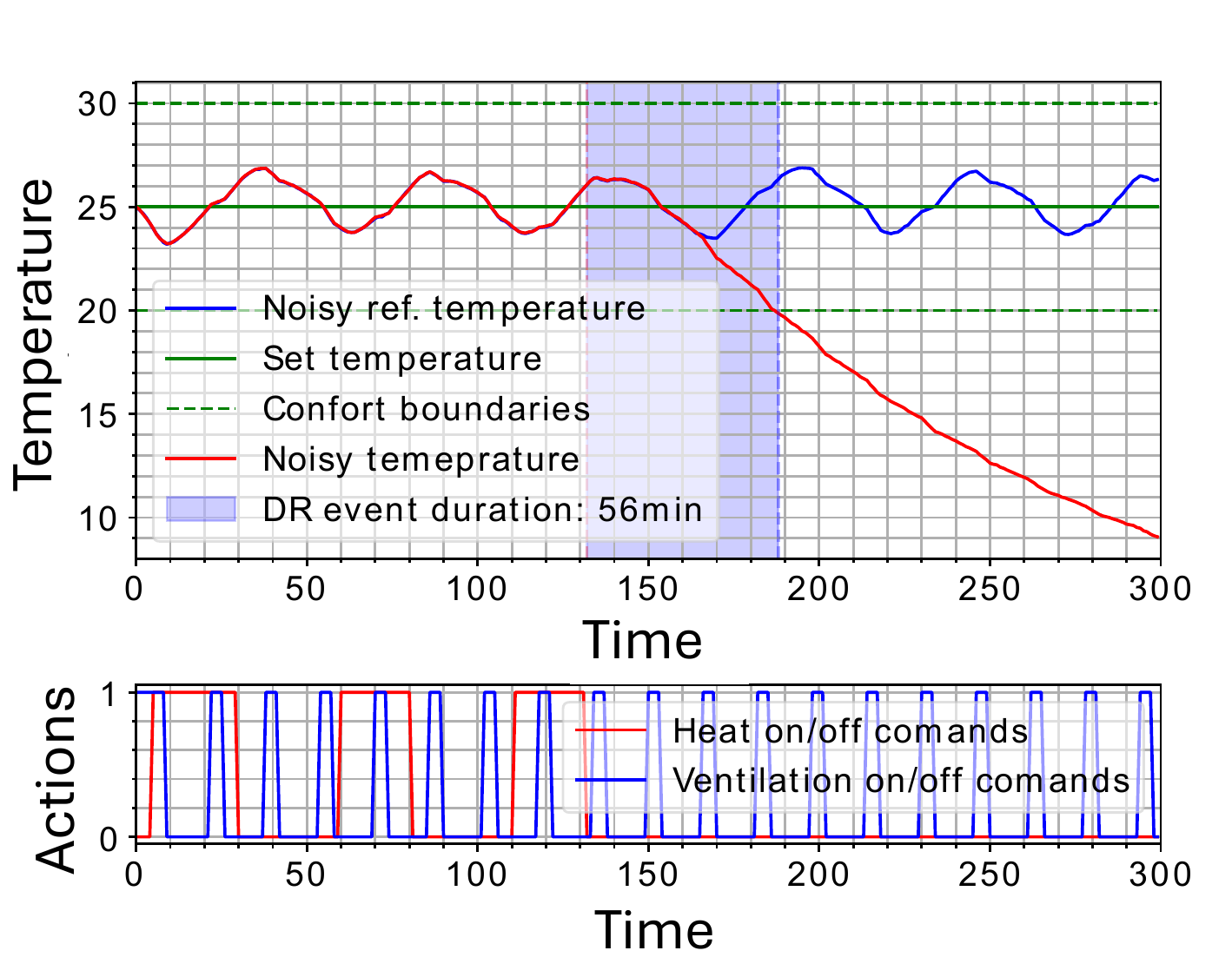}
  \caption{}
  \label{fig:sub2}
\end{subfigure}
\caption{Examples of planning and forecasting DR events with an outside temperature of 0 degrees Celsius and a comfort boundary of 20$^{\circ}$ C. \ref{fig:sub1} is the least desirable case, the DR event starts when the building is already close to the comfort boundary, and the maximum duration of the DR event is 26 minutes. \ref{fig:sub2} is the most desirable scenario achievable with planning and forecasting, the DR  event starts in phase with the end of the heating cycle, maximizing the time the DR event can last respecting the comfort boundary, the duration of the planned event is 56 minutes. }
\label{fig:planning}
\end{figure*}

Here, we give the example of a greenhouse that produces tomatoes in northern Europe with an HVAC system targeted at 25$^{\circ}$ C with a dead zone of $\pm$ 1$^{\circ}$ C.The thermal comfort zone of Tomatoes is well-known, and the effects of steeping out of it have also well-studied consequences \cite{BAZGAOU2021237}. For instance, a heated greenhouse with tomatoes requires keeping the temperature between 22$^{\circ}$ C and 26$^{\circ}$ C during the day and between 13$^{\circ}$ C and 18$^{\circ}$ C during the night \cite{BAZGAOU2021237}. A controlled anticipated wind-down regime could be part of a DR participation strategy, reducing the temperature down to 20$^{\circ}$ C or directly to 18$^{\circ}$ C for the night regime. It is known that the indoor agriculture industry suffers from high energy costs, so participating in DR using their thermal load could yield significant economic benefits \cite{PENUELA2024123756}. As greenhouses have a large thermal mass they can participate longer than most buildings, making them a perfect fit for thermal load DR. The principal task for a planner using our model would be to calculate how long the greenhouse can participate in each event. Our model is the perfect fit for calculating the length of the DR event. Given that our model takes as input the commands of the HVAC system, it is possible to use it to predict the behaviour of the room temperature in case of participation in a demand response program. Forecasting and planning DR events can highly impact their efficiency and the savings obtained by both the energy system and the user. When we participate in a DR event using a thermal load, we aim to maximise the length of the event, as most HVAC equipment operates only at nominal power and regulates the microclimate manipulating the duty cycle (ON/OFF state ratio). In Fig.~\ref{fig:planning}  we show an example of the best-case scenario and the worst-case scenario for our simulated greenhouse, showing the impact of planning and forecasting the DR event. A planned event can last more than two times longer than an unplanned event.

To fine-tune the model, experimental data is required, turning off the HVAC system and recording the changes until the temperature goes out of the comfort zone. We simulated such an experiment for five different temperatures for summer and five more for winter and augmented the dataset by moving the event in time, in other words sliding the predicting window back and forward in time. Also, we included different comfort thresholds where we calculated the prediction error. As we are predicting for this task the moment in time when the temperature drops or rises beyond the comfort limit, the best metric is RMSE. A comparison of the results of the model before and after fine-tuning is shown in Table~\ref{tab:DR_finetuned}. Note that as the data is generated using a minute-by-minute simulation, our model is predicting the length of the DR event as integers, which increases significantly the error of the model since a prediction deviation from the real value of 0.1 minute results in a 1 minute predicted deviation. For practical purposes, the model cannot use higher temporal resolution data, as the weather data current minimum resolution is one minute. Note that, as of yet, there is no benchmark that allows us to test the performance of our approach for demand response. However, works associated to DR, while not directly comparable to the present research due to differences in methodology and objectives, show that this is an active area of research \cite{VINDEL2023112686, GEHBAUER2023113481}.

\begin{table}
    \centering
    \begin{tabular}{|l|lcc|} 
    \hline
           &Metric&General model&  Fine-tuned model\\ 
           \cline{2-4}
           Cooling&RMSE&21.96& 
    2.944\\ 
  &Mean & 5.706&$-$1.824\\ 
  &std& 12.43&2.3\\ 
             \hline
 & Metric& General model&Fine-tuned model
\\            \cline{2-4}
 Heating& RMSE& 35.79&3.59
\\ 
 & Mean & 7.456&$-$1.342 
\\ 
 & std& 13.118&3.105\\ 
  \hline
 \end{tabular}

    \caption{Results of the  DR event length prediction on the generated 5 experimental data points (500 after argumentation) for fine-tuning, all the recorded values are in minutes. }
    \label{tab:DR_finetuned}
\end{table}

\subsubsection{Fault detection}
\label{sec:Faults}
As our model accounts for outside weather dynamics and the commands from the HVAC system, the model is consistent with the weather and the installed HVAC capacity. The model will not be sensitive to changes in the real HVAC system, in other words, the model can be used as a benchmark for correct HVAC behaviour and to diagnose the state of the HVAC system. To illustrate this possible application, once again, we can use our toy example of an industrial greenhouse where the yield of the plants can be severely affected if the greenhouse spends a few hours outside of the temperature comfort zone. In fact, if a failure in the HVAC system goes undetected during an extreme weather event, even the whole production can be lost. For instance, during the night regime in tomato production, a greenhouse in winter time, that loses its heating could fall from the lower comfort boundary of 12$^{\circ}$ C to temperatures close to zero, and the plants would be damaged by frostbite, with possible total losses.  Also, in less severe cases fault detection could still be useful. For instance, if the greenhouse is participating in DR, delayed fault detection in the HVAC system could lead to unplanned backup system use during a DR event and economic loss due to fines. Industrial greenhouses are usually located a few hours away from the city, so unplanned transport of equipment and qualified workers is problematic. Early failure detection could lead to some palliative measures to extend the time window for fixing the HVAC and reducing or eliminating vegetable production losses. In our example, we explore the following types of faults:
\begin{enumerate}
    \item The heating is off and physically unresponsive, but the electronic system is still responsive. 
    \item The heating is on and physically unresponsive, but the electronic system is still responsive. 
    \item The cooling is off and physically unresponsive, but the electronic system is still responsive. 
    \item The cooling is on and physically unresponsive, but the electronic system is still responsive.
    \item The thermal envelope is suddenly broken, equivalent to thermal insulation degradation.
\end{enumerate}

Using our model as a fault detector presents some limitations. First, our model can be used to reliably predict sensor degradation only when the degradation represent noise levels larger than $\pm 2.0 ^\circ $C; for lower noise levels our model's robustness to measurement noise is such that the changes in noise levels would be filtered out. Also, the model can estimate the correct functioning of the HVAC equipment or the thermal envelope only when the disruption affects the behaviour of the building. For instance, in the unlikely event that a window is broken when the building temperature is already thermalized with the environment, no change would be observed. 

To test the efficiency of the general model (i.e. without fine-tuning for this specific application) fault detection capabilities in the HVAC system we performed a simulation of 500 faulty events for each fault type and 500 non-faulty events. Using this dataset we performed a binary classification task, classifying an event as faulty or not for each fault case separately. We selected the RMSE at a 10-minute prediction horizon as the classification threshold which allows us to achieve close to 100\% precision and recall scores. The results are presented in Table \ref{tab:Faults}. Note that, we report the prediction time for correct classification, as we found a speed-precision trade-off. When compared to the state-of-the-art models for fault detection in HVAC systems, our model outperforms all reviewed models.  Hu et al. achieved an F1 score of 80\% \cite{HU2019117}, whereas our lowest F1 score (for fault type 3) was 98\%.  Zhang et al., in their meta-study, compared several fault detection models, reporting sensitivity (recall) values ranging from 0.86 up to 0.996 \cite{ZHANG2023100235}, while our model achieves 1.0 in most fault cases, and averages a sensitivity of 0.9972 for all cases. As industries still rely on expert knowledge for fault detection and diagnostics, models that facilitate prior knowledge integration are well-suited to meet industrial needs \cite{HEIMARANDERSEN2024113801}.

\begin{table}
    \centering
    \begin{tabular}{|c|cccc|}
\hline
 Fault &\multicolumn{4}{c|}{Best metrics}\\
 \cline{2-5}
          \# &  RMSE&  Recall&  Precision & Mean time
(Minutes)\\ \hline
         1
&  0.8&  0.992&   1& 13.67\\
         2
&  0.8&  1&   1& 14.03\\
         3
&  0.8&  0.994&   0.967& 7.25\\
         4
&  0.8&  1&   0.988& 7.38\\
         5&  0.8&  1&    0.983&  9.20$^*$
\\
         \hline

    \end{tabular}
    \caption{Results of the classification of fault events. $^*$ Due to the nature of the event, we could only approximate ground true values.}
    \label{tab:Faults}
\end{table}

\section{Validation using open-source datasets}
In this section we analyze the conditions that define an optimal dataset for full model implementation and testing. We selected three publicly available datasets for model testing and implementation recommendations: PLEIAD \cite{antonio_martinez_ibarra_2022_7620136}; LBNLL FDD \cite{OEDI_Dataset_5763}; and IATOBC \cite{soufi_2023_10000773}. These three datasets are described further below.

\subsection{Optimal dataset composition}
The optimal dataset for model implementation should contain:
\begin{itemize}
    \item  $T_{\rm out}$: High-resolution, high-fidelity outdoor temperature measurements, with 1-minute or 5-minute sampling data. The data should be smooth to ensure gradient stability. Data fidelity should be better than $T_{\rm obs}$, which is the measured indoor temperature. Poor quality weather data would prevent the deep Kalman filter to learn noise parameters.
    \item  $T_{\rm obs}$: High-resolution, 1-minute or 5-minute data, with data fidelity not lower than $2^\circ$C. The data should be smooth to ensure gradient stability. 
    \item $a_{\rm h}$, $a_{\rm vent}$, $a_{\rm ac}$: The heating, ventilation and cooling state variables (control commands) should be discrete state (preferably binary: ON or OFF) and represent the instantaneous state of the HVAC system (not daily, weekly or monthly HVAC modes). If the variables are discrete but not binary, all possible states should be less than 16 as the model uses a one-hot wrapping of depth 4. As the model is pretrained on binary states, using a different representation would require more data.
    \item Dataset span: The data set should represent a full year dynamics, including information such as, e.g., seasonality, energy saving setpoints, occupancy. A building in a climate with strong seasonal changes will require at least a year of data. A building with low seasonal variability could be trained on at least three months of data. 
\end{itemize}

For proper experimental validation the dataset should have indoor temperature data from two different sensors -- one high-quality and one low-quality sensor; this permits a straightforward comparison between the denoised output of the model and the high-quality sensor measurements. Without this type of measurement, the coefficient of determination metric becomes less meaningful. The dynamics prediction part of the model is trained on denoised data (as all data passes through the deep Kalman filter before prediction), meaning that it generates smooth temperature vs time predictions. Note that a smooth prediction curve could be a good approximation of the modeled data, and still have a negative coefficient of determination as one can compare only the predicted smooth output with the noisy data. 

\subsection{Dataset description}
\label{sec:validation}
All datasets were preprocessed by resampling to 1-minute measurements using interpolation for temperatures and nearest neighbor for the HVAC commands, converting temperatures in Fahrenheit to Celsius, filtering  non-differentiable shapes using the ScyiPy Savitzky–Golay filter. The HVAC commands were reconstructed from different control data provided in the datasets. Unfortunately, none of the datasets provide adequate control data for our model.  

A short description of each dataset follows:

\paragraph{PLEIAD dataset} This dataset presents a one-year fully curated dataset collected in  multi-zone buildings, including occupancy, at the University of Murcia in Spain \cite{antonio_martinez_ibarra_2022_7620136}. The dataset has high fidelity measurements of the outdoor temperature and indoor temperature. However, the sampling resolution is low (10 minutes). The HVAC control information represents control modes, not the actual control commands. The HVAC energy consumption is aggregated for the whole building, so actual HVAC use is not recoverable from the energy data. 
    
\paragraph{LBNL FDD dataset FCU} This is an EnergyPlus single-year single-room simulated dataset for fault detection \cite{OEDI_Dataset_5763}; note that we used the benchmark faultless dataset. The simulated building has occupancy hours, with sudden changes at the beginning and the end of the occupancy hours. The full control data is provided and a binary control dataset is derived from the continuous control data.  

\paragraph{Indoor Air Temperature and Occupant Behavior in Classroom (IATOBC) of higher education building in Mediterranean climate dataset} This is a three-month dataset collected in a multi-zone building with multiple sensors in the selected room \cite{soufi_2023_10000773}. The outdoor temperature measurements are noisier than the indoor temperature measurements. We approximated the HVAC control from window and door opening information and the heating/cooling hourly working mode. No information is provided on the HVAC mode (heating or cooling).

\begin{figure*}
\hspace*{-2.25cm} 
\centering
\begin{subfigure}{0.65\textwidth}
  \centering
  \includegraphics[width=\linewidth]{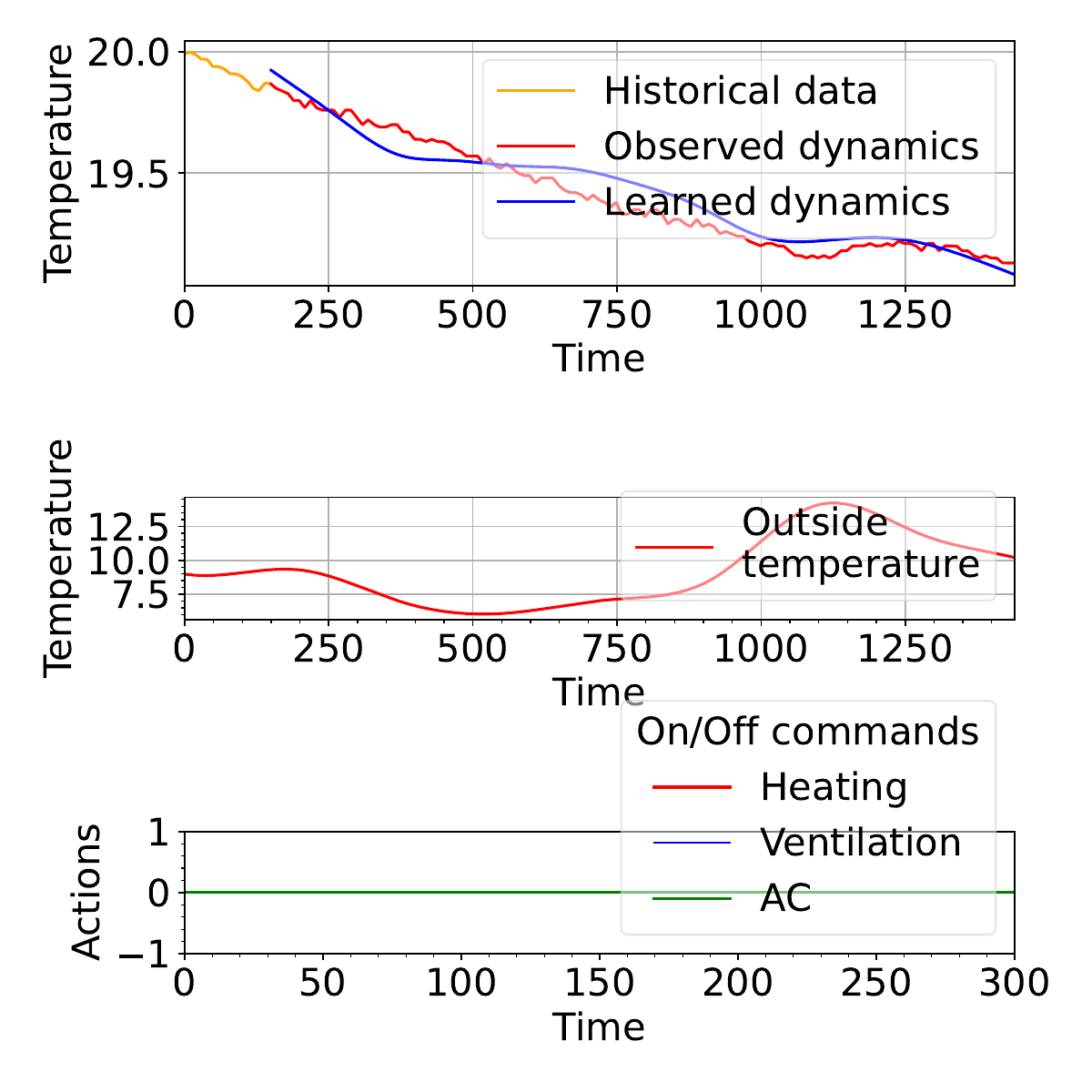}
  \caption{}
  \label{fig:val_sub1}
\end{subfigure}%
\begin{subfigure}{0.65\textwidth}
  \centering
  \includegraphics[width=\linewidth]{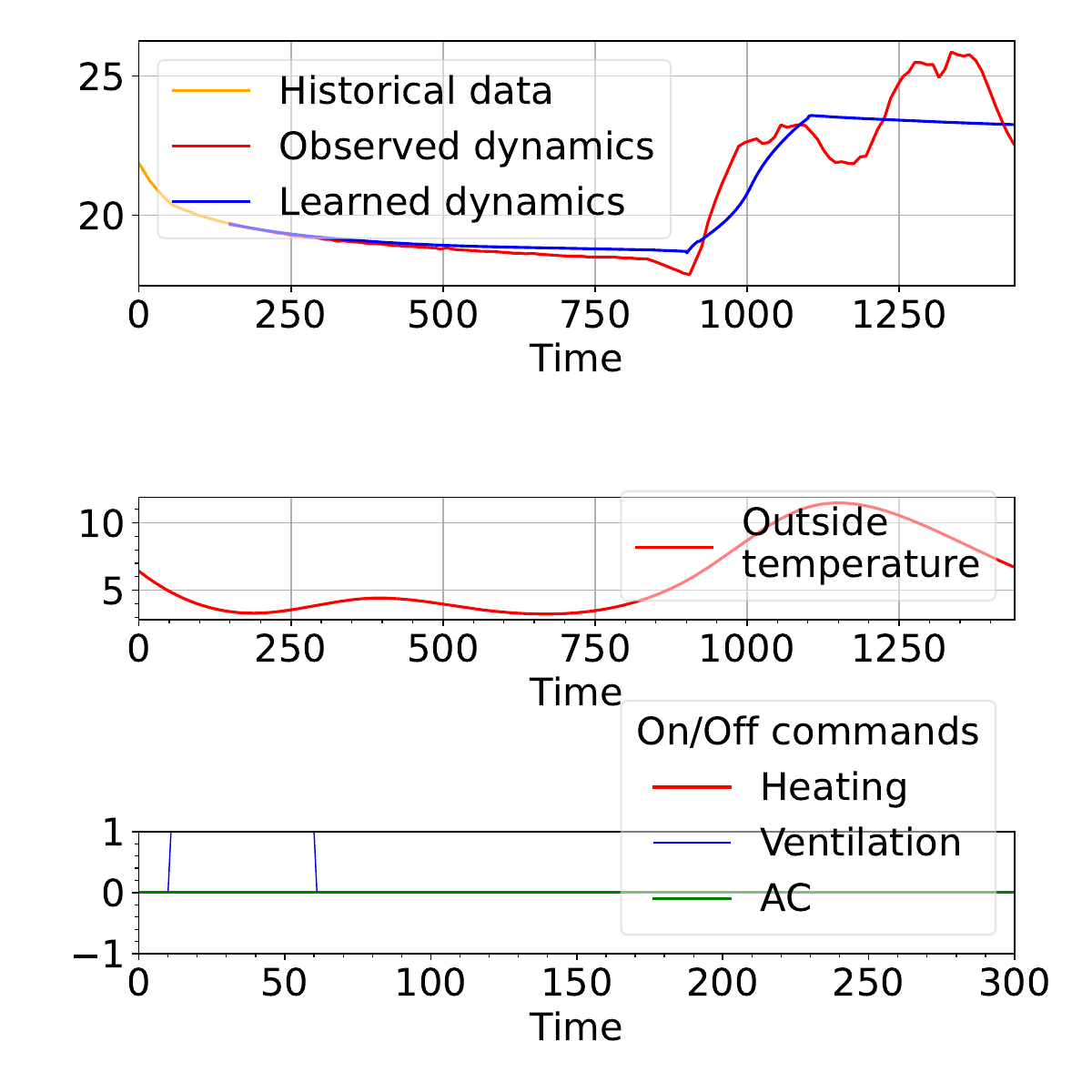}
  \caption{}
  \label{fig:val_sub2}
\end{subfigure}
\caption{Validation test samples: (\subref{fig:val_sub1}) best performing sample from the IATOBC dataset. (\subref{fig:val_sub2}) Sample from the IATOBC dataset showing oversmoothing in the model output.}
\label{fig:validation}
\end{figure*}

\begin{figure}
    \centering
    \includegraphics[width=\linewidth]{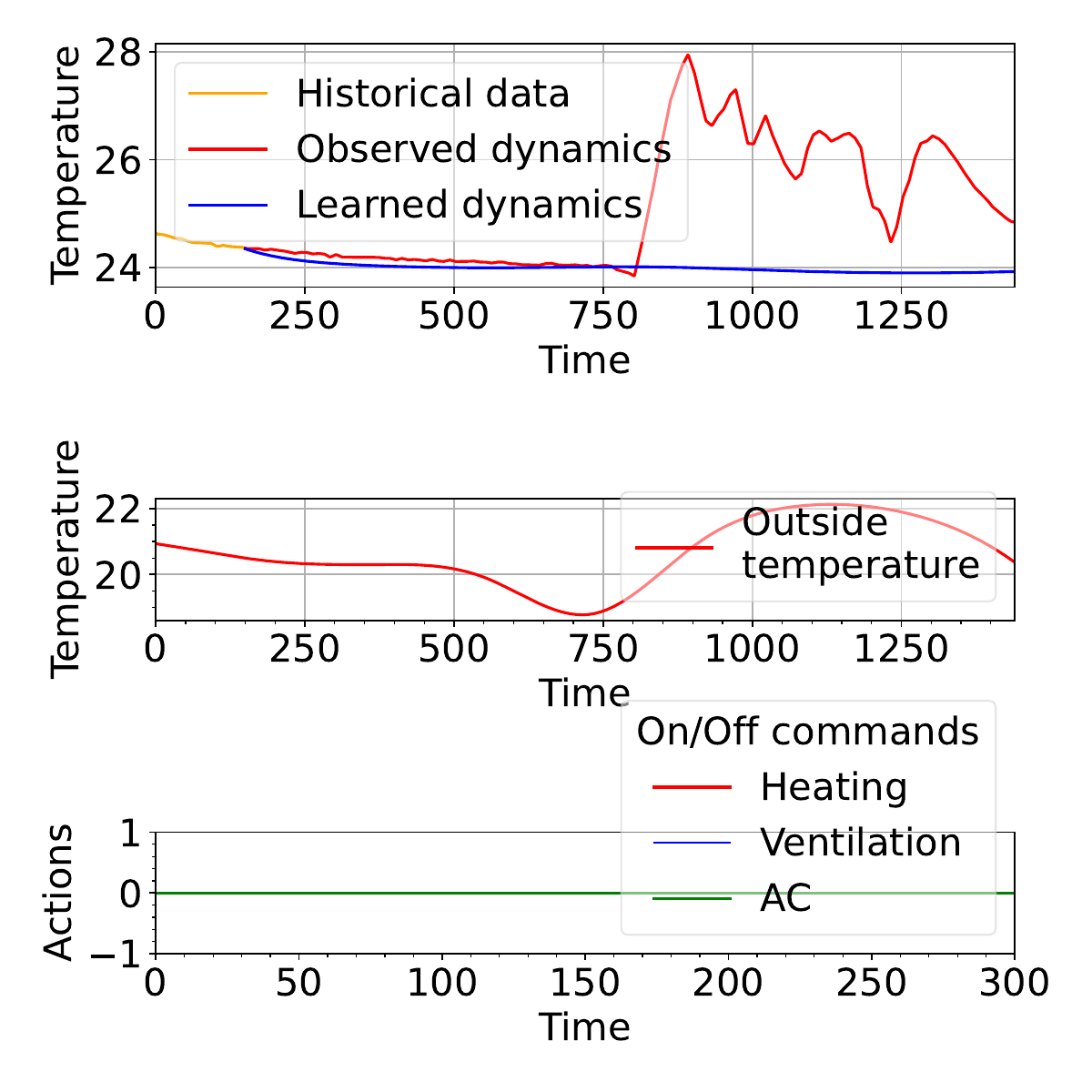}
    \caption{Validation test sample from the IATOBC dataset with a visible tendency for the model to reject changes that deviate from an expected setpoint.}
    \label{fig:val_constant}
\end{figure}

Given the focus on technological accessibility this type of model has, we limit the use of computational resources for finetuning to an equivalent of freely accessible GPU usage (about 4.5 hours in Google Colab). Concretely, this represents about 5000 training steps using batches of 4096 randomized samples. The results of the finetuning of our model to the each of the above dataset are presented in Table \ref{tab: valdation}. The resulting low performance when measured by the coefficient of determination $R^2$ is expected as we compare denoised predictions to noisy data. A coefficient of determination close to 1 would mean that our model overfits the data and does not perform denoising and prediction on denoised data. A visual analysis shows that the model has a tendency to smooth sharp changes, sometimes overshooting. We provide an example of the model predicted indoor temperature in Fig.~\ref{fig:validation}. On the left side, we see that the model is performing well while on as shown on the right side, oversmoothing causes loss of information. We also note that in the absence of HVAC control, external weather conditions are the main driver of the indoor temperature prediction. In Fig.~\ref{fig:val_constant}, we can see that the model expects the room to be controlled and has a bias towards keeping specific set temperatures showing that our initial training dataset is biased towards fully controlled environments. In our initial training dataset, lack of HVAC commands together with slow changes in outdoor temperature are seen only when the room is close to thermalization with the environment inside of the dead band. 

\begin{table}
    \centering
    \begin{tabular}{|c|c|c|c|c|}\hline
         \textbf{Dataset}&\textbf{RMSE}&  \textbf{MSE}&  \textbf{MAE} & \textbf{$R^{2}$}\\\hline\hline
         \textbf{PLEIAD}&0.575& 0.3904 & 0.462 & -1.07 \\ \hline
         \textbf{LBNL FDD data set FCU}&0.204&  0.0890&  0.174& -6.356\\\hline
         \textbf{IATOBC}&0.1758&  0.0449&  0.153& -25.46\\\hline
    \end{tabular}
    \caption{Model best performance on a 150-step (minutes) prediction horizon for the different available datasets after 5000 training steps. All metrics were calculated against noisy dataset.}
    \label{tab: valdation}
\end{table}

\subsection{Implementation recommendations}
During the validation tests the model necessitated the satisfaction of several requirements for robust finetuning. The first one is to have proper HVAC control sequences. The PLEIAD dataset has rooms with different dynamics; so, logically, the rooms where we see the most control activity (switching on and off the different HVAC elements) are the ones that leads the best finetuning. This is due to the changes of HVAC state having a higher correlation with the temperature trends; indeed if the HVAC system is constantly on or constantly off, the changes in the room temperature are not correlated to the HVAC system control commands. Another requirement is interpreting the sequence lengths as a hyperparameter. Our generated dataset is very ``eventful'': in a 300-minute sequence, there are several cycles of heating and cooling. The validation datasets on the other hand show less event, and require larger sequences of 1440 minutes or more to expose the model to full cycles. Increasing the sequence length from 300 to 1440 minutes during the finetuning stage, yields significantly better performance. If the amount of data is insufficient to successfully finetune the model, a user could tweak our initial simulation to generate a dataset that resembles the target building and train the model on it as a preliminary step to finetuning. 

\section{Conclusion}
In this work, we developed two methods for indoor temperature modeling. One is based on a simple step-by-step description of the environment with an iterative approach similar to what can be obtained from the solutions of a mathematical model that would describe this environment. This was used to generate synthetic data. The other is a statistics-based model where the complex probability density functions and their parameters are found using artificial neural networks and trained using the Bayesian approach. The principles of Kalman filtering were implemented together with a predictive model with memory, capable of accounting for thermal inertia. The training method is based on variational inference, which is one of the most efficient probabilistic approaches to machine learning \cite{igea_chatzis_cicirello_2022}. 

The resulting model architecture embodies both a denoiser and a predictor which can be used separately, giving the user the ability to integrate the denoiser and predictor contiguously or separately in different parts of the control loop. We demonstrated that its predicting capabilities correspond to the state-of-the-art models \cite{delcroix2021autoregressive,fang2021multi,AFROZ201829}, with the added advantage that it is deployable independently of the sensing hardware, due to its robustness to measurement noise. As our model can be used to control other physical quantities, including relative humidity, battery voltage, and more, and is applicable to various use cases beyond control systems, we evaluated it over different time horizons. The results showed peak performance for time horizons of a few hours. The model can be fine-tuned for use in different buildings without significant performance losses. Additionally, two use cases for the model, other than its implementation in indoor temperature control, are explored: demand response with thermal loads and HVAC fault detection, along with an example of industrial applications for both cases. Rapidly developing industries such as data-centers and other AI development infrastructure, in particular, can benefit from our proposed model. 

Although our model can be fine-tuned with relatively small datasets, it must be noted it only contains prior knowledge about thermal inertia and measurement noise. Consequently, the dataset must still capture the full range of temperature dynamics as it cannot infer the physical parameters governing the heat flow in the modeled building. Yet, the requirements for the dataset are much smaller: a heavily pruned dataset that keeps a few hours of each month could be sufficient to fine-tune the model. We explore the matter in the validation section where using datasets of 3-month and 1-year data, our model yields acceptable performance, even though the datasets do not provide sufficient data on HVAC commands. Next, the adaptability of the deep Kalman filter may conceal existing problems with the sensors, preventing timely solutions. Importantly, however, Bae et al. showed that 45\% of experts consider that the biggest barrier for widespread sensor implementation in indoor microclimate management is the initial cost, and 39\% consider that accuracy is also an issue \cite{BAE2021100068}. Our approach addresses those concerns. Our model has the advantage of allowing the deployment of low-cost sensors, with lower accuracy, or the use of already existing sensors, while getting performance comparable to the state-of-the-art high-quality sensors. Our work aims to foster technological accessibility for buildings with outdated HVAC sensing systems and control technologies by developing a model well-suited for MPC that could enhance performance of the control system. However, when narrow bandwidth connectivity or privacy are the primary concerns, models like the one presented by Zhenan Feng et al. \cite{FENG2024111385} could be a better choice. Given that the computational performance of servers is limited by thermal management, developments like our model are highly beneficial for such industries by potentially bringing down costs while increasing servers performance \cite{GARIMELLA201366,challe6010117,Corcoran_2013}.

\section*{Acknowledgments} J. P., S. M. H., I. K., and H. O. acknowledge partial support by the Skoltech program: Skolkovo Institute of Science and Technology -- Hamad Bin Khalifa University Joint Projects. The authors thank Dr. Alexander Ryzhov for useful discussions during the early stage of the work and a critical reading of the manuscript.


\newpage
\bibliographystyle{elsarticle-num}
\bibliography{bibliography}
\newpage
\appendix
\section{Simulated single-thermal-zone building description}
\label{Appendix:A}
Here, we present the specifics of the simulated single-thermal-zone building as a continuation of the description provided in Section \ref{simulation}. The simulation is implemented as a function that updates, step by step, all the values of the HVAC control state based on the previous step's indoor and outdoor temperatures, set temperature, and dead-band parameters. Using these updated values, the function calculates an update to Equations \ref{eq:Tobs}-\ref{eq:Tout}. For intuitive understanding, this approach can be thought of as a piecewise linear approximation (most functions are linear, although some are nonlinear) of the overall dynamics.

It is important to note that the control systems take measurement noise into account. As a result, noisier measured temperatures lead to more chaotic control system behavior. This intentional inclusion of noise provides a more robust test for our model and enhances its generalization capabilities. While using the observed temperature after denoising would improve our model’s performance, it would come at the cost of real-life applicability. Denoised data for fine-tuning the model on real-world buildings would require expensive high-end measurement equipment over months prior to deployment, which would defeat our purpose, as it demands the acquisition of such equipment. The deep Kalman filter integrated into our model is robust to a wide range of noise levels. After fine-tuning, our model allows for denoising without prediction, thereby simplifying HVAC control, so no sacrifice in usability is made.

The user is required to input the initial state of the system (initial set of observed and unobserved temperatures), the number of minutes for the simulation, the standard deviation of the noise level, the set temperature, the ventilation levels to maintain, and the size of the dead zone for the HVAC control. Based on the outside temperature, the system selects either the cooling or heating regime for the next 5 hours of simulation. Consequently, a single simulation may include switching between cooling and heating if the outside temperature necessitates it. Within the dead-band, no updates are made to the HVAC state. Outside of this range, the HVAC will turn the heating or cooling on or off as needed. Ventilation operates analogously. Once the states of the HVAC control system are defined, the initial state for the observed temperatures and the HVAC control is saved. The building's temperatures (including non-observed temperatures), the outside temperature (which can either be constant or modeled as a realistic weather simulation), and the degree of ventilation are updated using a single function. To prevent extreme parameter values, boundaries are implemented within the function. Gaussian noise is added to the indoor temperature, and the loop is repeated for each minute of the simulated environment. The pseudocode illustrating the entire process is provided in Algorithm \ref{alg:controlled_session}, while the pseudocode for the temperature update function is detailed in Algorithm \ref{alg:step}. The complete set of equations defining the observed and unobserved states of the simulated building is presented in Equations \ref{eq:A.1}-\ref{eq:A.7}.

The full code, along with the dataset, is publicly available in the GitHub repository at \url{https://github.com/Javier-ppp}

\begin{algorithm}
\caption{Simulation: general procedure}\label{alg:controlled_session}
\begin{algorithmic}[1]
\Procedure{Simulation}{n, $\sigma$, $t_{\text{set}}$, $v_{\text{set}}$, $\Delta t$, $\Delta v$} $\gets$\text{\it{simulation length(one step=one minute),std of measurement noise,}}
\text{\it{ desired temperature and ventilation values, and dead-band size}} $(\pm \Delta t, \Delta v)$.

    \State Initialize vectors for environment parameters and actions history collection.
    \State Set initial values: $t, t_{\text{out}},t_{\text{noisy}}, a, HoC$  $\gets$\text{\it{indoor temperature,}}
    \text{\it{outside temperature, indoor temperature with measurement noise,}} \text{\it{binary coded set of HVAC states, flag for heating or cooling regime.}}

    \For{$i \gets 0$ to $n - 1$}
        \If{$i \bmod 300 = 0$}
            \State $HoC \gets (t_{\text{out}} < t_{\text{set}} - \Delta t)$
        \EndIf

        \If{$t_{\text{noisy}} \geq t_{\text{set}} + \Delta t$}
            \State $a[0] \gets 0$ if $HoC$, else $a[2] \gets 1$
        \ElsIf{$t_{\text{noisy}} \leq t_{\text{set}} - \Delta t$}
            \State $a[0] \gets 1$ if $HoC$, else $a[2] \gets 0$
        \EndIf

        \If{$self.vent\_degree > v_{\text{set}} + \Delta v$}
            \State $a[1] \gets 0$
        \ElsIf{$self.vent\_degree < v_{\text{set}} - \Delta v$}
            \State $a[1] \gets 1$
        \EndIf

        \State Store $t$, $t_{\text{noisy}}$, $t_{\text{out}}$ and $a$ in respective collections
        \State Scale step $i$ to a 1440 min range (for $\cos$ sampling)
        \State Update $t$, $t_{\text{out}}$ using $stepfunction(a, \text{scaled step})$
        \State Add Gaussian noise to $t$ for $t_{\text{noisy}}$
    \EndFor

    \State \textbf{return} collections of $t$, $t_{\text{noisy}}$, actions, and $t_{\text{out}}$
\EndProcedure
\end{algorithmic}
\end{algorithm}

\begin{algorithm}
\caption{Simulation: Step function}\label{alg:step}
\begin{algorithmic}[1]
\Procedure{Stepfunction}{a, scaled step}
    \State Adjust $scaled\ step$ to within $[0, \pi]$ 
    \State Update temperature variables according to equations \ref{eq:A.1}-\ref{eq:A.5}

    \State Update $t_{\text{out}}$:
    \If{$c = 1$} 
        \State $t_{\text{out}} \gets t_{\text{out}}$ 
    \Else
        \State $t_{\text{out}} \gets $Eq. \ref{eq:A.7}
    \EndIf

    \State Enforce boundaries for $t_h$:
    \If{$t_h < 273.15$}
        \State $t_h \gets 273.15$
    \ElsIf{$t_h > 323.15$}
        \State $t_h \gets 323.15$
    \EndIf

    \State Update and enforce boundaries for $vent\_degree$:
    \State $vent\_degree \gets Eq.$ \ref{eq:A.6}
    \If{$vent\_degree < 0$}
        \State $vent\_degree \gets 0$
    \ElsIf{$vent\_degree > 60$}
        \State $vent\_degree \gets 60$
    \EndIf

    \State \textbf{return} $t$, $vent\_degree$, $t_{\text{out}}$
\EndProcedure
\end{algorithmic}
\end{algorithm}

\begin{eqnarray}
T_{\rm obs}^{\rm t+1} &=& T^{\rm t+1} + \epsilon\\
\nonumber
\label{eq:A.1}
T^{\rm t+1} &=&T^{\rm t}+ (T_{\rm h}^{\rm t} - T^{\rm t})\cdot 0.032  + (T_{\rm w}^{\rm t} - T^{\rm t})\cdot 0.025\\ 
&+& a_{\rm vent}^{\rm t} (T_{\rm out}^{\rm t}- T^{\rm t}) \cdot 0.005 + a_{\rm ac}^{\rm t}  (T_{\rm ac}^{\rm t}- T^{\rm t}) \cdot 0.016\\
T_{\rm h}^{\rm t+1} &=&T_{\rm h}^{\rm t}+ (T_{\rm fluid}^{\rm t}-T_{\rm h}^{\rm t})( a_{\rm h}^{\rm t}\cdot0.08) + (T^{\rm t} - T_{\rm h}^{\rm t})\cdot 0.05\\
\label{eq:A.5}
T_{\rm w}^{\rm t+1} &=& T_{\rm w}^{\rm t}+(T^{\rm t} - T_{\rm w}^{\rm t})\cdot0.1 + (T_{\rm out}^{\rm t} - T_{\rm w}^{\rm t})\cdot0.1\\
\label{eq:A.6}
V^{\rm t+1} &=& V^{\rm t} + a_{\rm vent}^{\rm t} \rm \cdot3 - (1 - a_{\rm vent}^{\rm t})\rm \cdot 1\\
\nonumber
T_{\rm out}^{\rm t+1} &=&T_{\rm out}^{\rm t} + (T_{\rm out}^{\rm 1}-T_{\rm out}^{\rm t})\cdot 0.001 +\cos(f(t)-0.7)\cdot 0.01\\ 
&+& X \cdot 0.01
\label{eq:A.7}
\end{eqnarray}
where $T_{\rm ac}^{\rm t}=const=283.15^\circ C$, $T_{\rm fluid}^{\rm t}$ is the linear interpolation from the outside temperature $T_{\rm out}^{\rm t}$ in the range $[253.15,318.1]$ to the heater working fluid temperature range $[340.15,298.15]$. Note that as the changes in $T_{\rm out}$ are smooth the changes in $T_{\rm fluid}$ are smooth too. $f(t)$ is the rounded (to the closest smallest number) normalized time step bounded to $[0,1]$ multiplied by $\pi$.  $X$ is a random value sampled from the distribution $\mathcal{N}(0,1)$

\end{document}